\documentclass[11pt,draftcls,onecolumn]{IEEEtran}

\usepackage{amsmath,amsfonts,amssymb,amsthm}
\usepackage{mathrsfs}
\usepackage{comment,blkarray}
\usepackage{multirow,bigdelim}
\usepackage{cite}
\usepackage[ruled,commentsnumbered, vlined]{algorithm2e}
\usepackage{tikz}
\usetikzlibrary{arrows,automata}
\usepackage[latin1]{inputenc}
\usepackage{verbatim}
\usepackage{graphicx}
\usepackage{cases}
\usepackage{booktabs}
\usepackage{subcaption}	
\usepackage{etoolbox}

\usepackage{enumitem}
\usepackage{ifpdf}															%
\ifpdf																		%
	\usepackage{hyperref}
\else																		%
\fi	
\usetikzlibrary{arrows,decorations.pathmorphing,decorations.footprints,fadings,calc,
trees,mindmap,shadows,decorations.text,patterns,positioning,shapes,matrix,fit}

\makeatletter
\newcommand{\rmnum}[1]{\romannumeral #1}
\newcommand{\Rmnum}[1]{\expandafter\@slowromancap\romannumeral #1@}

\newif\if@borderstar
\def\bordermatrix{\@ifnextchar*{%
  \@borderstartrue\@bordermatrix@i}{\@borderstarfalse\@bordermatrix@i*}%
}
\def\@bordermatrix@i*{\@ifnextchar[{\@bordermatrix@ii}{\@bordermatrix@ii[()]}}
\def\@bordermatrix@ii[#1]#2{%
\begingroup
  \m@th\@tempdima8.75\p@\setbox\z@\vbox{%
    \def\cr{\crcr\noalign{\kern 2\p@\global\let\cr\endline }}%
    \ialign {$##$\hfil\kern 2\p@\kern\@tempdima & \thinspace %
    \hfil $##$\hfil && \quad\hfil $##$\hfil\crcr\omit\strut %
    \hfil\crcr\noalign{\kern -\baselineskip}#2\crcr\omit %
    \strut\cr}}%
  \setbox\tw@\vbox{\unvcopy\z@\global\setbox\@ne\lastbox}%
  \setbox\tw@\hbox{\unhbox\@ne\unskip\global\setbox\@ne\lastbox}%
  \setbox\tw@\hbox{%
    $\kern\wd\@ne\kern -\@tempdima\left\@firstoftwo#1%
    \if@borderstar\kern2pt\else\kern -\wd\@ne\fi%
    \global\setbox\@ne\vbox{\box\@ne\if@borderstar\else\kern 2\p@\fi}%
    \vcenter{\if@borderstar\else\kern -\ht\@ne\fi%
    \unvbox\z@\kern -\if@borderstar2\fi\baselineskip}%
    \if@borderstar\kern -2\@tempdima\kern2\p@\else\,\fi\right\@secondoftwo#1 $%
  }\null \;\vbox{\kern\ht\@ne\box\tw@}%
\endgroup
}
\makeatother

\newtheorem{thm}{Theorem}
\newtheorem{lemma}[thm]{Lemma}
\newtheorem{cor}[thm]{Corollary}

\newtheorem{prop}[thm]{Proposition}
\newtheorem{example}{Example}
\newtheorem{defn}{Definition}
\allowdisplaybreaks

\newcommand{\eSET}{{\mathrm{E\text{-}SET}}}
\newcommand{\nSET}{{\mathrm{N\text{-}SET}}}

\newcommand{\tail}{{\mathrm{tail}}}
\newcommand{\head}{{\mathrm{head}}}
\newcommand{\Out}{{\mathrm{Out}}}
\newcommand{\In}{{\mathrm{In}}}
\newcommand{\mincut}{{\mathrm{mincut}}}

\newcommand{\w}{{\omega}}
\newcommand{\p}{{\rho}}
\newcommand{\bx}{{\bf x}}
\newcommand{\bz}{{\bf z}}
\newcommand{\tby}{{\tilde{\bf y}}}
\newcommand{\bzero}{{\bf 0}}
\newcommand{\de}{{\delta}}
\newcommand{\dt}{{\delta}}

\newcommand{\e}{\hat{e}}

\newcommand{\f}{\tilde{f}}

\newcommand{\y}{\tilde{y}}
\newcommand{\F}{\widetilde{F}}
\newcommand{\tC}{\widetilde{C}}

\newcommand{\mA}{\mathscr{A}}

\newcommand{\mB}{\mathscr{B}}
\newcommand{\mH}{\mathscr{H}}

\newcommand{\mF}{\mathbb{F}_q}
\newcommand{\Fq}{\mathbb{F}_q}

\newcommand{\mO}{\mathcal{O}}
\newcommand{\mE}{\mathcal{E}}
\newcommand{\msE}{\mathscr{E}}

\newcommand{\mC}{\mathcal{C}}
\newcommand{\mZ}{\mathcal{Z}}
\newcommand{\Rank}{{\mathrm{Rank}}}

\newcommand{\row}{{\rm row}}

\newcommand{\wtG}{\widetilde{G}}

\SetKw{KwInitialization}{Initialization:}
\allowdisplaybreaks

\begin{document}

\title{Linear Network Error Correction Coding:\\
A Revisit
}

\author{Xuan~Guang~~and~~Raymond~W.~Yeung
\thanks{This paper was presented in part at the 2020 IEEE International Symposium on Information Theory.}
}
\date{}
\maketitle

\begin{abstract}
We consider linear network error correction (LNEC) coding when errors may occur on edges of a communication network of which the topology is known. In this paper, we first revisit and explore the framework of LNEC coding, and then unify two well-known LNEC coding approaches. Furthermore, by developing a graph-theoretic approach to the framework of LNEC coding, we obtain a significantly enhanced characterization of the error correction capability of LNEC codes in terms of the minimum distances at the sink nodes. In LNEC coding, the minimum required field size for the existence of LNEC codes, in particular LNEC maximum distance separable (MDS) codes which are a type of most important optimal codes, is an open problem not only of theoretical interest but also of practical importance, because it is closely related to the implementation of the coding scheme in terms of computational complexity and storage requirement. By applying the graph-theoretic approach, we obtain an improved upper bound on the minimum required field size. The improvement over the existing results is in general significant. The improved upper bound, which is graph-theoretic, depends only on the network topology and requirement of the error correction capability but not on a specific code construction. However, this bound is not given in an explicit form. We thus develop an efficient algorithm that can compute the bound in linear time. In developing the upper bound and the efficient algorithm for computing this bound, various graph-theoretic concepts are introduced. These concepts appear to be of fundamental interest in graph theory and they may have further applications in graph theory and beyond.
\end{abstract}


\IEEEpeerreviewmaketitle

\section{Introduction}

In 1956, the problem of maximizing the rate of flow from a source node to a sink node through a network was considered independently by Elias~\textit{et al.}~\cite{Elias-Feinstein-Shannon-maxflow} and Ford and Fulkerson~\cite{maxflow-Ford-Fulkerson}, where, regardless of whether the flow is a commodity flow or an information flow, the value of the maximum flow is equal to the capacity of a minimum cut separating the sink node from the source node. This result is the celebrated {\em max-flow min-cut theorem}, proved in \cite{Elias-Feinstein-Shannon-maxflow} and \cite{maxflow-Ford-Fulkerson}.
In 2000, Ahlswede~\textit{et al.}~\cite{Ahlswede-Cai-Li-Yeung-2000} put forward the general concept of {\em network coding} that allows the intermediate nodes in a noiseless network to process the received information. In particular, they focused on the single-source network coding problem on a general network and proved that if coding is applied at the nodes in a network, rather than routing only, the single source node can multicast messages to all the sink nodes at the theoretically maximum rate, i.e., the smallest minimum cut capacity between the source node and a sink node, as the alphabet size of both the information source and the channel transmission symbol tends to infinity. This result can be regarded as the max-flow min-cut theorem for information flow from a source node multicasting to multiple sink nodes through a network, as well as a generalization of the classical max-flow min-cut theorem from a source node to a sink node through a network. The idea of network coding can be dated back to Celebiler and Stette's work \cite{Celebiler-Stette-1978} in 1978, where they proposed a scheme that can improve the efficiency of a two-way satellite communication system by performing the addition of two bits onboard the satellite. In 1999, Yeung and Zhang \cite{Zhang-Yeung-1999} investigated the general coding problem in a satellite communication system and obtained an inner bound and an outer bound on the capacity region.
Shortly after \cite{Elias-Feinstein-Shannon-maxflow}, Li~\textit{et~al.} \cite{Li-Yeung-Cai-2003} proved that linear network coding with a finite alphabet is sufficient for optimal multicast by means of a vector space approach. Independently, Koetter and M\'{e}dard \cite{Koetter-Medard-algebraic} developed an algebraic characterization of linear network coding by means of a matrix approach. The above two approaches correspond to the {\em global} and {\em local} descriptions of linear network coding, respectively. For comprehensive discussions of network coding, we refer the reader to \cite{Zhang-book, Yeung-book, Fragouli-book, Fragouli-book-app, Ho-book}.

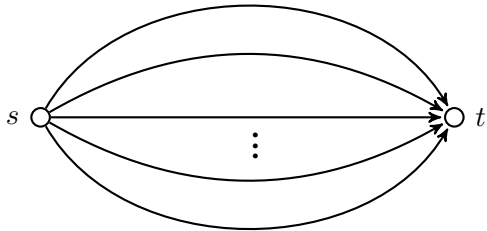
\begin{figure}[!t]
\centering
\begin{tikzpicture}
[->,>=stealth',shorten >=1pt,auto,node distance=5cm, thick]
  \tikzstyle{every state}=[fill=none,draw=black,text=black,minimum size=7pt,inner sep=0pt]
  \node[state]         (s)[label=left:$s$]                 {};
  \node[state]         (t)[right of=s, xshift=5mm, label=right:$t$] {};
\path
(s)   edge[bend left=60]       node[pos=0.5, right=-0.5mm]{}(t)
      edge[bend left=30]       node[pos=0.5, right=-0.5mm]{}(t)
      edge                     node[pos=0.52, below=-1.5mm]{\bf$\vdots$}(t)
      edge[bend right=30]      node[pos=0.5, right=-0.5mm]{}(t)
      edge[bend right=60]      node[pos=0.5, right=-0.5mm]{}(t);
\end{tikzpicture}
\caption{An equivalent model of the classical coding theory.}
\label{Fig_classical_coding_model}
\end{figure}

In the paradigm of network coding, network error correction is necessary when errors may occur on the edges of a communication network. For example, network transmission may suffer from random errors caused by channel (edge in networks) noise, erasure errors caused by link failure or buffer overflow, corruption errors caused by malicious attack, etc. In general, the problem induced by errors in network coding can be more serious than the one in a classical point-to-point communication system, because errors will be propagated by the coding operations at the intermediate nodes. Even a single error occurred on an edge has the potential of polluting all the ``downstream'' messages. The network coding techniques for combating network errors is referred to as {\em network error correction coding}. In particular, the linear network coding techniques for combating network errors is referred to as {\em linear network error correction (LNEC) coding}, which was introduced in~\cite{Yeung-Cai-coorrect} and investigated widely in the literature, e.g., \cite{Yeung-Cai-correct-1,Yeung-Cai-correct-2,Yang-refined-Singleton,zhang-correction,Koetter-correction,
Silva-K-K-rank-metric-codes,Guang-MDS,Zhang-survey-paper-NEC,Guang-Zhang-NECBook}.
A very special case of network error correction coding over the simplest network is depicted in Fig.~\ref{Fig_classical_coding_model}, where the network consists of only two nodes, a source node $s$ and a sink node $t$, connected by multiple parallel edges from $s$ to $t$. This special case of network error correction coding can be regarded as the model of classical coding theory (cf.~\cite{MacWilliams-Sloane-Theory-error-correting-codes_book, Huffman-Pless-Fundamentals-error-correting-codes_book}), which is a very rich field of research originated from Shannon's seminal work \cite{Shannon48} in 1948.

\subsection{Related Works}

Network error correction coding was first considered by Cai and Yeung \cite{Yeung-Cai-coorrect}. Subsequently, they further developed network error correction coding in their two-part paper \cite{Yeung-Cai-correct-1,Yeung-Cai-correct-2} as a generalization of algebraic coding from the point-to-point setting to the network setting. In particular, three important bounds in algebraic coding, the Hamming bound, the Gilbert-Varshamov bound, and the Singleton bound, are generalized for network error correction coding, where the error correction capabilities at all the sink nodes are the same. Subsequently, the Singleton bound was refined independently by Zhang~\cite{zhang-correction} and Yang~\textit{et al.}~\cite{Yang-refined-Singleton}, where the error correction capabilities at the sink nodes can be different. This refined Singleton bound shows that sink nodes with larger maximum flow values from the source node can have potentially higher error correction capability. Similar refinements for the Hamming bound and the Gilbert-Varshamov bound were also provided in~\cite{Yang-refined-Singleton}. In the rest of the paper, the refined Singleton bound will be called the Singleton bound for network error correction coding.

Two frameworks of LNEC coding were developed in \cite{zhang-correction} and \cite{Yang-weight}. In order to characterize error correction capability of an LNEC code, Zhang~\cite{zhang-correction} directly defined a minimum distance at sink node by using the introduced concept of the rank of error pattern, which can be regarded as a ``measure'' of error pattern. Subsequently, Guang~\textit{et al.}~\cite{Guang-MDS} proved that this minimum distance can be obtained by using other measures of error pattern. Yang~\textit{et al.}~\cite{Yang-refined-Singleton} considered multiple weight measures on error vector occurred in the network to characterize error correction capability of an LNEC code. They further proved that these weight measures induce the same minimum weight decoder.
The construction of LNEC codes has been investigated in the literature. In \cite{Yang-refined-Singleton,Matsumoto-Singleton,Guang-MDS}, different constructions of LNEC maximum distance separable (MDS) codes were put forward, where LNEC MDS codes are a type of most important optimal codes that achieve the Singleton bound with equality. These constructions also imply the tightness of the Singleton bound. Besides, the construction in \cite{Guang-MDS} can also be applied to construct a general LNEC code with any admissible requirement of the rate and error correction capability, which includes LNEC MDS codes as a special case. Further, Guang~\textit{et al.}~\cite{Guang-uni-MDS} considered the problem of network error correction coding when the information rate changes over
time. To efficiently solve this problem, local-encoding-preserving LNEC coding was put forward, where a family of LNEC codes is called local-encoding-preserving if all the LNEC codes in this family share a common local encoding kernel at each intermediate node in the network. In order to achieve the maximum error correction capability for each possible rate, an efficient approach was also provided to construct a family of local-encoding-preserving LNEC MDS codes with all the admissible rates.

A common assumption in the above discussion is that the network topology is known. As such, we can construct a deterministic LNEC code based on the network topology, and use this code for network transmission. By contrast, for the case that the network topology is unavailable, it is impossible to construct an LNEC code based on the network topology. Network error correction coding without this assumption has been investigated in the literature. One approach is random LNEC coding \cite{zhang-random,Guang-MDS,Zhang-survey-paper-NEC,Cai,Guang-uni-MDS}, which uses the same idea in random network coding first studied by Ho~\textit{et al.}~\cite{Ho-etc-random}. To be specific, this approach applies random network coding to build the extended global encoding kernels for each sink node, which form a matrix for decoding the source message with error correction. Another approach is subspace coding \cite{Koetter-correction,Silva-K-K-rank-metric-codes,KSK-subspace-codes}, which is an end-to-end approach for error correction with random linear network coding employed within the network. To be specific, in this approach, random linear network coding over a network is abstracted as an {\em operator channel} in K\"{o}tter and Kschischang's work \cite{Koetter-correction}. The source node, as the transmitter of this operator channel, emits a vector space modulated by a source message. A sink node, as a receiver of this channel, receives a vector space which is possibly corrupted by network errors. A new metric, called {\em subspace distance}, is used to measure the discrepancy between the two vector spaces for network error correction. With this metric, efficient coding and decoding schemes based on rank-metric and subspace codes were proposed in \cite{Koetter-correction,Silva-K-K-rank-metric-codes,Silva-Kschischang-Metrics}.

Another line of research considers adversarial attacks, in which various adversarial models were investigated in the context of network coding \cite{Ho-Byzantine, Jaggi-Byzatine-IT08,Nutman-Langberg-Byzatine-ISIT08, Silva-Kschischang-Metrics,Kosut-Tong-Tse-Jaggi-Polytope_codes-IT14}. In particular, for the Byzantine attack in which an adversary is able to modify the messages transmitted on the edges of a network \cite{Ho-Byzantine, Nutman-Langberg-Byzatine-ISIT08, Jaggi-Byzatine-IT08}, network error correction coding can be applied to combat the attack by regarding the malicious messages injected into the network by the adversary as errors. For example, Jaggi~\textit{et al.}~\cite{Jaggi-Byzatine-IT08} proposed a distributed polynomial-time algorithm for correcting the corruption errors, which can achieve successful decoding with a high probability when the sizes of the base field and the source message packet are sufficiently large. A cryptographic technique for public-key systems is also used in their coding scheme. Specifically, a redundancy matrix, which plays the role of a parity-check, needs to be published in advance to all the parties including the source node, the sink nodes and the adversaries before employing (random) linear network coding within the network.

\subsection{Contributions and Organization of the Paper}

In this paper, we first revisit and further explore the framework of LNEC coding and network error correction on a network whose topology is known. Then, we show that the two well-known LNEC approaches developed in \cite{zhang-correction} and \cite{Yang-refined-Singleton} are in fact equivalent. By developing a graph-theoretic approach, we can enhance the characterization of error correction capability of LNEC codes in terms of the minimum distances at the sink nodes. Briefly speaking, in order to ensure that an LNEC code can correct up to $r$ errors at a sink node $t$, it suffices to ensure that this code can correct every error vector in a ``reduced set of error vectors''. In general, the size of this reduced set is considerably smaller than the number of error vectors with Hamming weight not larger than $r$. This result has the important implication that the computational complexities for decoding and code construction can be significantly reduced.

In LNEC coding, the minimum required field size for the existence of LNEC codes, in particular LNEC MDS codes, is an open problem not only of theoretical interest but also of practical importance, because it is closely related to the implementation of the coding scheme in terms of computational complexity and storage requirement~\cite{Yeung-Cai-correct-1,Yeung-Cai-correct-2,Yang-refined-Singleton,zhang-correction,Guang-MDS}. However, the existing upper bounds on the minimum required field size for the existence of LNEC (MDS) codes are typically too large for implementation. In this paper, we show that the required field size for the existence of LNEC (MDS) codes can be reduced significantly. To be specific, by applying our graph-theoretic approach, we prove an improved upper bound on the minimum required field size. The improvement over the existing results is in general significant. This new bound, which is graph-theoretic, depends on the network topology and the requirement of error correction capability but not on a specific code construction. As mentioned, our upper bound is graph-theoretic, but it is not given in an explicit form. Thus, we develop an efficient algorithm to compute the bound whose computational complexity is in a linear time of the number of edges in the network.

The paper is organized as follows. In Section~\ref{Sec_preliminaries}, we formally present the network model and linear network coding. The necessary notation and definitions are also introduced. In Section~\ref{sec:LNEC-revisited}, we revisit and explore the framework of LNEC coding, and then unify two well-known LNEC coding approaches. In Section~\ref{sec:enhanced_capability}, we develop a graph-theoretic approach with which we can enhance the characterization of error correction capability of LNEC codes. The improved upper bound on the minimum required field size for the existence of LNEC codes, in particular LNEC MDS codes, is obtained in Section~\ref{sec:field_size_reduction}. This is followed by the development of an efficient algorithm for computing the improved bound. We conclude in Section~\ref{sec:concl} with a summary of our results.

\section{Preliminaries}\label{Sec_preliminaries}

\subsection{Network Model}\label{Sec_II-A}

Let $G=(V,E)$ be a finite directed acyclic graph with a single source $s$ and a set of sink nodes $T\subseteq V\setminus \{s\}$, where $V$ and $E$ are the sets of nodes and edges, respectively. For a directed edge~$e$ from node $u$ to node $v$, the {\em tail} and the {\em head} of an edge $e\in E$ are denoted by $\tail(e)$ and $\head(e)$, respectively. Further, for a node $v$, let $\Out(v)=\{e\in E:\tail(e)=v\}$ and $\In(v)=\{e \in E:\head(e)=v\}$, which are the set of input edges and the set of output edges, respectively. Without loss of generality, assume that there are no input edges for the source node $s$ and no output edges for any sink node $t\in T$. The capacity of each edge is taken to be $1$, i.e., a symbol taken from an alphabet is transmitted on each edge $e \in E$ for each use of $e$. Further, parallel edges between two adjacent nodes are allowed.

In the network $G$, if a sequence of edges $(e_1,e_2,\cdots,e_m)$ satisfies $\tail(e_{k+1})=\head(e_k)$ for all $k=1,2,\cdots,m-1$, then $(e_1,e_2,\cdots,e_m)$ is called a {\em path} from the node $\tail(e_1)$ (or the edge $e_1$) to the node $\head(e_m)$ (or the edge $e_m$). In particular, a single edge $e$ is regarded as a path from $\tail(e)$ to $\head(e)$ (or from $e$ to itself). For two nodes $u$ and $v$, a {\em cut} separating $v$ from  $u$ is a set of edges whose removal disconnects $v$ from $u$, i.e., no paths exist from $u$ to $v$ upon deleting the edges in this set. The {\em capacity} of this cut separating $v$ from $u$ is defined as the number of edges in the cut. The minimum of the capacities of all cuts separating $v$ from $u$ is called the {\em minimum cut capacity} separating $v$ from $u$. Further, a cut is called a {\em minimum cut} separating $v$ from $u$ if its capacity achieves this minimum cut capacity. If $u$ and $v$ are two nodes such that $v$ is disconnected from $u$, i.e., no path exists from $u$ to $v$ in the network $G$, we adopt the convention that the minimum cut capacity separating $v$ from $u$ is $0$ and the empty set of edges is the minimum cut separating $v$ from $u$.

These concepts can be extended from separating a node $v$ from another node $u$ to separating a nonempty subset of nodes~$\widehat{V}$ from a node $u$ ($u\notin \widehat{V}$), and separating an edge subset $\xi$ from a node $u$ as follows. We first consider a nonempty subset of non-source nodes $\widehat{V}\subseteq V$. We create a new node $v_{\widehat{V}}$, and for every node $v$ in $\widehat{V}$, add a ``super-edge'' of infinite capacity from $v$ to $v_{\widehat{V}}$ (which is equivalent to adding an infinite number of parallel edges from $v$ to $v_{\widehat{V}}$). A {\em cut} separating $\widehat{V}$ from $u$ is defined as a cut of {\em finite} capacity separating $v_{\widehat{V}}$ from $u$. We can naturally extend the definitions of the capacity of a cut, the minimum cut capacity, and the minimum cut to the case of~$\widehat{V}$. Next, we consider an edge subset $\xi\subseteq E$. We first subdivide each edge $e\in \xi$ by creating a node $v_e$ and splitting $e$ into two edges $e^1$ and $e^2$ such that $\tail(e^1)=\tail(e)$, $\head(e^2)=\head(e)$, and $\head(e^1)=\tail(e^2)=v_e$. Let $V_\xi=\big\{v_e: e\in \xi \big\}$. Then a {\em cut} separating the edge subset $\xi$ from the node $u$ is defined as a cut separating $V_\xi$ from $u$, where, whenever $e^1$ or $e^2$ appears in the cut, replace it by~$e$. By definition, $\xi$ is a cut separating $\xi$ from $u$. Similarly, the minimum cut capacity separating $V_\xi$ from $u$ is defined as the \textit{minimum cut capacity} separating $\xi$ from $u$. Also, a cut separating $\xi$ from $u$ achieving this minimum cut capacity is called a \textit{minimum cut} separating $\xi$ from $u$. If an edge set $A\subseteq E$ is a cut separating a node $v$ (resp. a set of nodes $\widehat{V}$ and a set of edges $\xi$) from another node $u$, then we say that the edge set $A$ {\em separates} $v$ (resp.~$\widehat{V}$ and $\xi$) {\em from} $u$. Note that if $A$ separates $v$ (resp. $\widehat{V}$ and $\xi$) from $u$, then every path from $u$ to $v$ (resp. $\widehat{V}$ and $\xi$) passes through at least one edge in $A$. We now use a network in \cite{GYF-LEP-SNC} (Figs.~2 and~3) as an example to illustrate the above graph-theoretic concepts.

\begin{example}
\begin{figure}[t]
\tikzstyle{vertex}=[draw,circle,fill=gray!30,minimum size=6pt, inner sep=0pt]
\tikzstyle{vertex1}=[draw,circle,fill=gray!80,minimum size=6pt, inner sep=0pt]
\centering
\begin{minipage}[b]{0.5\textwidth}
\centering
{
 \begin{tikzpicture}[x=0.6cm]
    \draw (0,0) node[vertex]     (s)[label=above:$u$] {};
    \draw (-2,-1.3) node[vertex] (1)  {};
    \draw ( 2,-1.3) node[vertex] (2)  {};
    \draw ( 0,-2.6) node[vertex] (3)  {};
    \draw ( 0,-4) node[vertex] (6)  {};
    \draw (-2,-6) node[vertex] (4)  {};
    \draw ( 1.5,-5.5) node[vertex] (5)  {};

    \draw[->,>=latex] (s) -- (1) node[midway, left=-0.2mm] {$e_1$};
    \draw[->,>=latex] (s) -- (2) node[midway, auto, right=-0.2mm] {$e_2$};
    \draw[->,>=latex] (1) -- (3) node[midway, auto,left=-0.5mm] {$e_3$};
    \draw[->,>=latex] (2) -- (3) node[midway, auto,right=-0.5mm] {$e_4$};
    \draw[->,>=latex] (3) -- (6) node[midway, auto, left=-1mm] {$e_5$};
    \draw[->,>=latex] (6) -- (4) node[midway, auto, left=-0.5mm] {$e_{6}$};
    \draw[->,>=latex] (6) -- (5) node[midway, auto,swap, left=-0.5mm] {$e_{7}$};
    \draw[->,>=latex] (5) -- (4) node[midway, auto,swap, below=-0.2mm] {$e_{8}$};
    \end{tikzpicture}
}
\caption{The network $G$.}
\label{fig-network1}
\end{minipage}%
\centering
\begin{minipage}[b]{0.5\textwidth}
\centering
 \begin{tikzpicture}[x=0.6cm]
    \draw (0,0) node[vertex]     (s)[label=above:$u$] {};
    \draw (-2,-1.3) node[vertex] (1)  {};
    \draw ( 2,-1.3) node[vertex] (2)  {};
    \draw ( 0,-2.6) node[vertex] (3)  {};
    \draw ( 0,-3.3) node[vertex1] (7) [label=right:$v_{e_5}$] {};
    \draw ( 0,-4) node[vertex] (6)  {};
    \draw ( 0.8,-4.8) node[vertex1] (8) [label={[xshift=2mm, yshift=-2mm]$v_{e_7}$}] {};
    \draw (-2,-6) node[vertex] (4)  {};
    \draw ( 1.5,-5.5) node[vertex] (5)  {};
    \draw ( 4.5,-5.5) node[vertex1] (9) [label=below:$v_{\xi}$] {};

    \draw[->,>=latex] (s) -- (1) node[midway, left=-0.2mm] {$e_1$};
    \draw[->,>=latex] (s) -- (2) node[midway, auto, right=-0.2mm] {$e_2$};
    \draw[->,>=latex] (1) -- (3) node[midway, auto,left=-0.5mm] {$e_3$};
    \draw[->,>=latex] (2) -- (3) node[midway, auto,right=-0.5mm] {$e_4$};
    \draw[->,>=latex] (3) -- (7) node[midway, auto, left=0mm] {$e_5^1$};
    \draw[->,>=latex] (7) -- (6) node[midway, auto, left=0mm] {$e_5^2$};

    \draw[->,>=latex] (6) -- (4) node[midway, auto, left=-0.5mm] {$e_{6}$};
    \draw[->,>=latex] (6) -- (8) node[pos=0.7, left=-0.5mm] {$e_{7}^1$};
    \draw[->,>=latex] (8) -- (5) node[pos=0.7, left=-0.3mm] {$e_{7}^2$};
    \draw[->,>=latex] (5) -- (4) node[midway, auto,swap, below=-0.2mm] {$e_{8}$};

    \draw[->,>=latex,ultra thick, gray!70] (7) -- (9) node[pos=0.5, above=0.2mm] {\scriptsize{\color{black}$\infty$}};
     \draw[->,>=latex,ultra thick, gray!70] (8) -- (9) node[pos=0.5, below=0mm] {\scriptsize{\color{black}$\infty$}};
    \end{tikzpicture}
\caption{The network modification.}
\label{fig-network2}
\end{minipage}
\end{figure}

We consider node $u$ and an edge subset $\xi=\{e_5,e_7 \}$ in the network $G$ depicted in Fig.~\ref{fig-network1}. For edge $e_5$, we first create a node $v_{e_5}$ and split $e_5$ into two edges $e_5^1$ and $e_5^2$ with $\tail(e_5^1)=\tail(e_5)$, $\head(e_5^2)=\head(e_5)$, and $\head(e_5^1)=\tail(e_5^2)=v_{e_5}$. The same subdivision operation is applied to edge $e_7$ as depicted in Fig.~\ref{fig-network2}. Let $V_\xi=\big\{v_{e_5}, v_{e_7}\big\}$. Now, in order to find a cut separating $\xi$ from $u$, it is equivalent to finding a cut separating $V_\xi$ from $u$. Toward this end, we first create a new node $v_{\xi}$ and add 2 super-edges with infinite capacity from $v_{e_5}$ to $v_{\xi}$ and from $v_{e_7}$ to $v_{\xi}$, respectively. By definition, a cut of finite capacity separating $v_{\xi}$ from $u$ is a cut separating $V_{\xi}$ from $u$ and so a cut separating $\xi$ from $u$. For example, the edge subset $\{e_3,e_4\}$ is such a cut. Further, the edge subset $\{e_5^1\}$ is also a cut separating $V_{\xi}$ from $u$. By definition, $e_5^1$ appears in the cut $\{e_5^1\}$ and $e_5\in \xi$, and thus $e_5^1$ is replaced by $e_5$ and so $\{e_5\}$ is a cut separating $\xi$ from $s$. We further see that $\{e_5\}$ is a minimum cut separating $\xi$ from $u$ that achieves the minimum cut capacity $1$ separating $\xi$ from $u$.
\end{example}

Due to the acyclicity of the network $G$, we can fix an {\em ancestral order} on the edges in $E$ that is consistent with the natural partial order of the edges. Throughout this paper, we use this order to index the coordinates of all the vectors and the rows/columns of all the matrices in the paper. If the columns of a matrix $L$ are indexed by a subset of edges $\xi$, then we use a symbol with subscript $e$, say $\ell_e$, to denote the column indexed by the edge $e\in \xi$; 
if the rows of a matrix $L$ are indexed by the subset of edges $\xi$, then we use a symbol followed by $e$ in a pair of brackets, say $\ell(e)$, to denote the row indexed by $e\in \xi$.

\subsection{Linear Network Coding}\label{subsec_linear_network_coding}


In this subsection, we consider the linear network coding model. On the network $G$, the source node~$s$ is required to multicast the source message to each node in $T$, or equivalently, each node in $T$ is required to decode with zero error the source message generated by the source node $s$. For a sink node~$t\in T$, we use $C_t$ to denote the minimum cut capacity separating $t$ from the source node $s$. Linear network coding over a finite field is sufficient for achieving $\min_{t\in T}C_t$, the theoretical maximum rate at which the source node $s$ can multicast the source message to all the sink nodes in~$T$ \cite{Li-Yeung-Cai-2003, Koetter-Medard-algebraic}.

Let $\w$ be the {\em (information) rate} of the source ($\w\leq \min_{t\in T}C_t$), or equivalently, the source node $s$ generates $\w$ symbols in an alphabet per unit time. To facilitate our discussion, we introduce $\w$ {\em imaginary source edges} connecting to $s$, denoted by $d_1',d_2',\cdots,d_\w'$, respectively, and let $\In(s)=\big\{d_1',d_2',\cdots,d_\w' \big\}$. As such, we assume that the $\w$ source symbols are transmitted to $s$ on the $\w$ imaginary source edges. Now, we state the definition of a linear network code.

\begin{defn}\label{def-LNC}
Let $\Fq$ be a finite field of order $q$, where $q$ is a prime power. An $\Fq$-valued rate-$\w$ linear network code $\mC$ on the network $G=(V,E)$ consists of an $\Fq$-valued $\lvert \In(v)\rvert \times \lvert \Out(v)\rvert$ matrix $K_v=[k_{d,e}]_{d\in \In(v), e\in \Out(v)}$ for each non-sink node $v$ in $V$, i.e.,
\begin{align*}
\mC=\big\{ K_v:~v\in V\setminus T \big\},
\end{align*}
where $K_v$ is called the local encoding kernel of $\mC$ at $v$, and $k_{d,e}\in \Fq$ is called the local encoding coefficient for the adjacent edge pair $(d,e)$.
\end{defn}

For a linear network code $\mC$, the local encoding kernels induce a column $\w$-vector $f_e$ for each edge $e$ in $E$, called the \textit{global encoding kernel} of~$e$, which can be calculated recursively according to the given ancestral order of edges in $E$ by
 \begin{align}\label{equ_f_e}
        f_e=\sum_{d\in \In (\tail(e))}k_{d,e}\cdot f_d,
 \end{align}
with the boundary condition that $f_{d_i'}$, $1 \leq i \leq \w$ form the standard basis of the vector space $\Fq^\w$. The set of global encoding kernels for all $e\in E$, i.e., $\big\{ f_e:~e\in E \big\}$, is also used to represent this linear network code $\mC$. However, we remark that a set of global encoding kernels $\big\{ f_e:~e\in E \big\}$ may correspond to more than one set of local encoding kernels $\big\{ K_v:~v\in V\setminus T \big\}$.

In using this rate-$\w$ linear network code $\mC$, let $\bx=\big(x_1~x_2~\cdots~x_\w \big) \in \Fq^{\w}$ be the row vector of $\w$ source symbols generated by the source node~$s$, which is called the {\em source message vector}, or simply the {\em source message}. Without loss of generality, we assume that $x_i$ is transmitted on the $i$th imaginary channel $d'_i$, $1\leq i \leq \w$. We use $y_e$ to denote the symbol transmitted on $e$, $\forall~e\in \In(s)\bigcup E$. With $y_{d_i'}=x_i$, $1\leq i \leq \w$, each $y_e$ for $e\in E$ can be calculated recursively according to the given ancestral order of edges in $E$ by the equation
\begin{align}\label{equ_y_e}
y_e=\sum_{d\in \In(\tail(e))}k_{d,e}\cdot y_d.
\end{align}
In fact, $y_e$ is a linear combination of the $\w$ source symbols $x_i$, $1\leq i\leq \w$, which can be seen as follows. First, it is readily seen that $y_{d'_i}= \mathbf{x} \cdot f_{d'_i}$ $(=x_{i})$, $1\leq i \leq \w$. Then it can be shown by induction via \eqref{equ_f_e} and \eqref{equ_y_e} that
\begin{align}
y_e= \mathbf{x} \cdot f_{e}, \quad \forall\,e\in E.
\end{align}

For each sink node $t\in T$, we define the matrix $F_t=\Big[f_e:~e\in \In(t)\Big]$. The sink node~$t$ can decode the source message vector with zero error if and only if $F_t$ is full rank, i.e., $\Rank\big(F_t\big)=\w$. We say that a rate-$\w$ linear network code $\mC$ is {\em decodable for $T$} if for each sink node $t \in T$, the rank of the matrix $F_t$ is equal to the rate $\w$ of the code, i.e., $\Rank\big(F_t\big)=\w$, $\forall~t\in T$.\footnote{When the set of sink nodes $T$ is clear from the context, we say that the linear network code $\mC$ is ``decodable'' instead of ``decodable for $T$'' for simplicity.} We refer the reader to \cite{Zhang-book, Yeung-book, Fragouli-book, Fragouli-book-app, Ho-book} for comprehensive discussions of linear network coding.

\section{Linear Network Error Correction Coding Revisited}\label{sec:LNEC-revisited}

\subsection{Linear Network Error Correction Coding}\label{subsect-LNEC}

In this subsection, we present the linear network error correction (LNEC) coding model. We first consider using an $\Fq$-valued rate-$\w$ linear network code $\mC$ on the network $G=(V,E)$ to multicast the source message to the sink nodes in $T$. When the symbol $y_e$ is transmitted on edge $e$, an error $z_e \in \Fq$ may occur.\footnote{If no error occurs on the edge $e$, then $z_e=0$.} As a result, the output of edge~$e$ becomes $\tilde{y}_e=y_e+z_e$. The error $z_e$ is treated as a message called the \textit{error message} on edge $e$. We write all the errors on the edges in $E$ as an $\Fq$-valued row $|E|$-vector $\bz=(z_e:~e\in E)$ and call $\bz$ the {\em error vector}.

To take into account of the effect of the errors on the network $G$, we can modify the linear network code $\mC$ to a rate-$\w$ LNEC code on $G$. Before describing the modification, we first present the {\em extended network} $\widetilde{G}=\big(\widetilde{V},\widetilde{E}\big)$ of $G$, which was introduced in~\cite{zhang-correction}. In the original network $G$, for each edge $e\in E$, we introduce an imaginary edge $e'$ such that $\head(e')=\tail(e)$, which is called the {\em imaginary error edge} for edge $e$. Similar to the source message generated by the source node $s$, we also assume that the error $z_e$ is transmitted to $\tail(e)$ through the imaginary error edge~$e'$. The original network $G$ together with all the imaginary error edges $e'$, $e\in E$ form the {\em extended network} of $G$ denoted by $\widetilde{G}=(\widetilde{V},\widetilde{E})$, where $\widetilde{V}=V$ and $\widetilde{E}=E\bigcup E'$ with $E' \triangleq \big\{e':~e\in E \big\}$, the set of all the imaginary error edges. 
Clearly, the extended network $\widetilde{G}$ is also acyclic due to the acyclicity of the original network $G$. As for linear network coding, we introduce $\w$ imaginary source edges $d_1'$, $d_2'$, $\cdots$, $d_\w'$ connecting to the source node $s$ in the extended network $\widetilde{G}$, where $\w$ is the rate of the source, and let $\In(s)=\{d_1',d_2',\cdots,d_\w'\}$. For every non-source node $v$ on $\widetilde{G}$, we use $\In(v)$ to denote the set of ``real'' input edges of $v$, i.e., the imaginary error edges connected to $v$ are not included in $\In(v)$. Now, we modify the rate-$\w$ linear network code $\mC$ on $G$ into a rate-$\w$ linear network code on $\widetilde{G}$ by setting the local encoding coefficients with respect to each imaginary error edge $e'\in E'$ as follows:
\begin{align}\label{k_e'_d}
k_{e',d}=\left\{
  \begin{array}{ll}
    1, & \hbox{$d=e$;} \\
    0, & \hbox{$d \in \Out\big(\tail(e)\big)\setminus \{e\}$.}
  \end{array}
\right.
\end{align}
This modified linear network code on $\widetilde{G}$ is called the corresponding $\Fq$-valued rate-$\w$ LNEC code on the original network $G$. In the following, we define the global encoding kernels of such a rate-$\w$ LNEC code on $G$ in terms of the local encoding coefficients.

\begin{defn}\label{def-LENC}
Let $\mF$ be a finite field of order $q$, where $q$ is a prime power. An $\mF$-valued rate-$\w$ LNEC code on the network $G=(V,E)$ consists of a column $(\w+|E|)$-vector $\f_e$ for each edge $e$ in $E$, called the extended global encoding kernel of $e$, whose components are indexed by the $\w$ imaginary source edges in $\In(s)$ and the $|E|$ imaginary error edges in $E'$, such that
\begin{enumerate}
        \item $\f_{d_i'}=1_{d_i'}$, $1\leq i \leq \w$, $\f_{e'}=1_{e'}$, $e'\in E'$, form the standard basis of the vector space $\mF^{\w+|E|}$, where $1_{d}$, $d\in \In(s) \bigcup E'$ is a column $(\w+|E|)$-vector whose  component indexed by $d$ is equal to $1$ while all other components are equal to $0$;
        \item For each edge $e\in E$, $\f_e$ is calculated recursively according to the given ancestral order of edges in $E$ by
        \begin{align}\label{equ_ext_f}
        \f_e=\sum_{d\in \In(\tail(e))}k_{d,e}\cdot\f_d+1_{e'},
        \end{align}
        where $k_{d,e}\in \mF$ is the local encoding coefficient for the adjacent edge pair $(d,e)$.
\end{enumerate}
\end{defn}

In using this rate-$\w$ LNEC code on $G$, let $\bx=\big(x_1~x_2~\cdots~x_\w \big)$ be the source message vector and $\bz=(z_e:~e\in E)$ be the error vector. For each imaginary source edge $d_i'$, $1\leq i \leq \w$ and each imaginary error edge $e'\in E'$, we have, respectively,
\begin{align*}
\tilde{y}_{d_i'}=x_i \quad \text{ and } \quad \tilde{y}_{e'}=z_e.
\end{align*}
The symbol $\tilde{y}_{e}$, the output of edge $e\in E$, is recursively calculated by
\begin{align}\label{tilde_y_e}
\tilde{y}_{e}=\sum_{d\in \In(\tail(e))}k_{d,e}\cdot\tilde{y}_{d}+z_e
\end{align}
according to the given ancestral order of edges in $E$.
Comparing~\eqref{equ_ext_f} with~\eqref{tilde_y_e}, we obtain that
\begin{align}\label{tilde_y_f_e}
\tilde{y}_{e}=(\bx~\bz)\cdot \f_e,\quad \forall~e\in \In(s)\bigcup\widetilde{E}.
\end{align}

Before discussing how to use this LNEC code to correct errors on the network, we first introduce some notation to be used frequently throughout the paper. For an edge $e\in \widetilde{E}$, we write $\f_e$ as
\begin{equation}\label{tilde_f_e}
\f_e=\begin{pmatrix}
\f_e(d_1')& \cdots & \f_e(d_\w')& \f_e(e'_1)& \cdots& \f_e(e'_{|E|}) \end{pmatrix}^\top
=\begin{bmatrix}
f_e \\
g_e
\end{bmatrix},
\end{equation}
where
\begin{align}\label{f_e_g_e}
f_e=\begin{pmatrix}
\f_e(d_1')& \f_e(d_2')&
\cdots&
\f_e(d_\w')
\end{pmatrix}^\top
\quad
\text{and} \quad
g_e=\begin{pmatrix}
\f_e(e'_1)& \f_e(e'_2)&
\cdots&
\f_e(e'_{|E|})
\end{pmatrix}^\top.
\end{align}
Further, for a sink node $t \in T$, we let $\F_t=\begin{bmatrix}\f_e:\ e\in \In(t)\end{bmatrix}$, an $(\w+|E|)\times |\In(t)|$ matrix, and use $\row_{t}(d')$ to denote the row vector of $\F_t$ indexed by the imaginary edge $d' \in \In(s)\bigcup E'$, i.e.,
$\row_t(d')=\big( \f_{\e}(d'):~ \e \in \In(t) \big)$.
 Then, we write
\begin{align}\label{tilde_F_xi}
\F_t=
\begin{bmatrix}
F_t\\
G_t
\end{bmatrix},
\end{align}
where
\begin{align}\label{F_xi-G_xi}
F_t=\begin{bmatrix}
\row_{t}(d_1')\\
\vdots\\
\row_{t}(d_\w')
\end{bmatrix} \quad \text{and}\quad
G_t=\begin{bmatrix}
\row_{t}(e'_1)\\
\vdots \\
\row_{t}(e'_{|E|})
\end{bmatrix}
\end{align}
are two matrices of sizes $\w\times |\In(t)|$ and $|E|\times |\In(t)|$, respectively.

\subsection{Network Error Correction}\label{subsec:network-error-correction}

We consider an $\Fq$-valued rate-$\w$ LNEC code $\tC=\big\{\f_e:~e\in E \big\}$ on the network $G=(V, E)$. We first assume that $\tC$ is decodable for the set of sink nodes~$T$, i.e., $\Rank\big(F_t\big)=\w$, $\forall~t\in T$. Herein, the decodability property is necessary, because otherwise, even if no errors occur on the network, at least one of the sink nodes in $T$ cannot decode the source message with zero error.

Let $\bz=(z_e:~e\in E)\in \Fq^{|E|}$ be an error vector and $\p\subseteq E$ be an edge subset. We say that $\bz$ {\em matches}~$\p$ if $z_e=0$ for all $e\in E\setminus \p$, i.e.,
\begin{align}\label{equ_match}
\bz\in \Big\{\bz'=(z_e':~e\in E)\in \Fq^{|E|}:~ z_e'=0,~ \forall~e\in E\setminus \p \Big\}.
\end{align}
For notational convenience, we write~\eqref{equ_match} as $\bz\in \p$ in the rest of the paper. This abuse of notation should cause no ambiguity and would greatly simplify the notation.

We now consider network error correction. We assume that a sink node $t$ knows the extended global encoding kernels of the input edges of $t$, i.e., $\F_t$. For a source message vector $\bx\in \Fq^\w$ on $d_i'$, $1\leq i \leq \w$ and an error vector $\bz\in \Fq^{|E|}$ on $e'\in E'$, we denote by $\y_e(\bx, \bz)$ the symbol transmitted on an edge $e$. 
Further, we let
$$\tby_t(\bx,\bz) \triangleq \big(\tilde{y}_e(\bx,\bz):~e\in \In(t) \big),$$
and by~\eqref{tilde_y_f_e}, we have
\begin{align}\label{deccoding-equation-t}
\tby_t(\bx,\bz)=(\bx~\bz)\cdot\F_t.
\end{align}
When $\bx$ and $\bz$ are clear from the context, we write $\y_e$ and $\tby_t$ to simplify the notations.

At the sink node $t$, the source message vector $\bx$ and error vector $\bz$ are unknown while $\F_t$ and $\tby_t$ are known. We attempt to decode $\bx$ by ``solving'' $\bx$ in the equation $\tby_t=(\bx~\bz)\cdot\F_t$ in which $\bx$ and $\bz$ are regarded as variables.

We let $\mZ$ be a set of error vectors. We say that the rate-$\w$ LNEC code $\tC$ {\em corrects any error vector in~$\mZ$ at the sink node $t$} if for any 2 pairs $(\bx~\bz)$ and $(\bx'~\bz')$ such that $\tby_t(\bx,\bz)=\tby_t(\bx',\bz')$, where $\bx, \bx'\in \mF^\w$ and $\bz, \bz' \in \mZ$, we have
\begin{align*}
\bx=\bx'.
\end{align*}
As such, we see that any source message vector $\bx\in \Fq^\w$ can be decoded with zero error regardless which error vector in $\mZ$ occurs in the network.

Next, we consider the error correction capability of an $\Fq$-valued rate-$\w$ LNEC code $\tC=\big\{\f_e:~e\in E \big\}$ on the network $G=(V, E)$, i.e, the possible set of error vectors for each sink node $t\in T$ in which any error vector can be corrected by $\tC$ at $t$. We first define two types of vector spaces for the code $\tC$, which play a crucial role for network error correction \cite{zhang-correction,Guang-MDS,Guang-Zhang-NECBook}.

\begin{defn}\label{defn_min_dis}
Consider a sink node $t\in T$ and an edge subset $\p \subseteq E$. At the sink node $t$, the message space and the error space of $\p$ are defined, respectively, by
\begin{align}\label{equ-source_error_spaces}
\Phi(t)=\big\langle \row_t(d_i'):\ 1\leq i\leq \w \big\rangle\quad
\text{and}\quad
\Delta(t,\p)=\big\langle \row_t(e'):\ e\in \p \big\rangle.\footnotemark
\end{align}
\end{defn}
\footnotetext{Here we use $\big\langle L \big\rangle$ to denote the subspace spanned by the vectors in a set $L$ of vectors.}

With Definition~\ref{defn_min_dis}, we readily see that
\begin{align}\label{Phi_xi}
\Phi(t)=\Big\{ \bx\cdot F_t:~\text{all source message vectors}~\bx\in \mF^{\w} \Big\},
\end{align}
and
\begin{align}\label{Delta_xi}
\Delta(t,\p)=\Big\{ \bz\cdot G_t:~\text{all error vectors~$\bz\in \mF^{|E|}$ such that $\bz \in \p$} \Big\}.
\end{align}
For a source message vector $\bx\in \mF^\w$ and an error vector $\bz \in \mF^{|E|}$ such that $\bz\in\p$, by~\eqref{tilde_y_f_e}, \eqref{tilde_f_e} and \eqref{f_e_g_e}, we have
\begin{align*}
\tilde{y}_e=(\bx~\bz)\cdot\f_e=\bx\cdot f_e + \bz\cdot g_e, \quad \forall~e\in E.
\end{align*}
By \eqref{tilde_F_xi} and \eqref{F_xi-G_xi}, we immediately have
\begin{align}\label{tby_t}
\tby_t =(\bx~\bz)\cdot\F_t=\bx\cdot F_t+ \bz\cdot G_t.
\end{align}
Thus, we observe that the ``effect'' of $\bx$ (i.e., $\bx\cdot F_t$) at $t$ belongs to $\Phi(t)$ by~\eqref{Phi_xi} and the ``effect'' of $\bz \in \p$ (i.e., $\bz\cdot G_t$) at $t$ belongs to $\Delta(t,\p)$ by~\eqref{Delta_xi}. Briefly speaking, if the ``effect'' $\bz\cdot G_t$ of the error vector $\bz$ at $t$ can be removed from $\tby_t$, then, together with $\Rank(F_t)=\w$, the source message vector $\bx$ can be decoded with zero error. This will become clear in the following discussions.

With the equation~\eqref{tby_t}, the ``effect'' $\bx\cdot F_t$ of a source message vector $\bx$ can be regarded its ``codeword'' at the sink node $t$, in which $F_t$ is regarded as the ``generator matrix'' at $t$. So $\Phi(t)$ can be regarded as the ``codebook'' at $t$. We now consider 2 different codewords $\bx\cdot F_t$ and $\bx'\cdot F_t$ (i.e., $\bx \neq \bx'$ by $\Rank(F_t)=\w$). Based on the above discussions, either $\bx$ or $\bx'$ cannot be decoded with zero error if and only if there exists 2 vectors $\bz$ and $\bz'$ such that $\tby_t(\bx, \bz)= \tby_t(\bx',\bz')$, or equivalently,
\begin{align}\label{equ4}
(\bx-\bx')\cdot F_t = (\bz'-\bz) \cdot G_t.
\end{align}
Hence, we define the distance between 2 codewords $\bx\cdot F_t$ and $\bx'\cdot F_t$ as follows:
\begin{align}
d^{(t)}(\bx\cdot F_t,~ \bx'\cdot F_t)=\min\Big\{ |\p|:~\exists \text{ an error vector } \bz\in\p \text{ s.t. } (\bx-\bx')\cdot F_t=\bz\cdot G_t \Big\}.
\end{align}
Before proving that $d^{(t)}(\cdot,\cdot)$ is a metric, we first extend the distance between 2 codewords to the distance between 2 vectors in $\Fq^{|\In(t)|}$.
\begin{defn}\label{defn_distance}
Consider a rate-$\w$ LNEC code $\tC$ on the network $G$ and a sink node $t\in T$. For any 2 vectors $\tby_t$ and $\tby_t'$ in $\Fq^{|\In(t)|}$, the distance between $\tby_t$ and $\tby_t'$ is defined as
\begin{align}\label{distance}
d^{(t)}(\tby_t, \tby_t')=\min\Big\{ |\p|:~\exists \text{ an error vector } \bz\in\p \text{ s.t. } \tby_t-\tby_t'=\bz\cdot G_t \Big\}.
\end{align}
\end{defn}

In~\eqref{distance}, when $\tby_t=\tby_t'$, the edge subset $\p$ that achieves the minimum is the empty set with  the error vector $\bz$ being the all-zero vector. By~\eqref{equ_ext_f} and \eqref{F_xi-G_xi}, we can obtain that the $|\In(t)|\times |\In(t)|$ submatrix $\begin{bmatrix}  \row_t(e'):~e\in \In(t) \end{bmatrix}$ of $G_t$ is an identity matrix (cf.~the proof of Theorem~\ref{lem_equi_Edt_Adt} in Section~\ref{sec:enhanced_capability} for more details). So for any 2 vectors $\tby_t$ and $\tby_t'$ in $\Fq^{|\In(t)|}$, there must exist an error vector $\bz$ such that $\tby_t-\tby_t'=\bz\cdot G_t$. Then, the distance $d^{(t)}(\cdot,\cdot)$ is well-defined.

\begin{prop}\label{prop_distance}
The distance $d^{(t)}(\cdot,\cdot)$ defined in the vector space $\Fq^{|\In(t)|}$ is a metric, i.e., the following 3 conditions are satisfied for arbitrary vectors $\tby_t$, $\tby_t'$ and $\tby_t''$ in $\Fq^{|\In(t)|}$:
  \begin{enumerate}
     \item {\bf(Positive Definiteness)}~ $d^{(t)}(\tby_t, \tby_t')\geq 0$, and $d^{(t)}(\tby_t, \tby_t')=0$ if and only if $\tby_t=\tby_t'$;
     \item {\bf(Symmetry)}~ $d^{(t)}(\tby_t, \tby_t')=d^{(t)}(\tby_t', \tby_t)$;
     \item {\bf(Triangle Inequality)}~ $d^{(t)}(\tby_t, \tby_t'')\leq d^{(t)}(\tby_t, \tby_t')+d^{(t)}(\tby_t', \tby_t'')$.
  \end{enumerate}
\end{prop}
\begin{IEEEproof}
See Appendix~\ref{pf_prop_distance}.
\end{IEEEproof}\medskip

Thus, the pair $\big(\Fq^{|\In(t)|}, d^{(t)}(\cdot,\cdot)\big)$ forms a metric space. Furthermore, we naturally define the {\em minimum distance} of the codebook $\Phi(t)$, denoted by $d_{\min}^{(t)}$, as
\begin{align*}
d_{\min}^{(t)}=\min_{\bx, \bx'\in \Fq^{\w}:~\atop \bx \neq \bx'} d^{(t)}\big(\bx\cdot F_t,~\bx'\cdot F_t\big).
\end{align*}

We continue to consider the distance between two codewords:
\begin{align}\label{equ5}
d^{(t)}(\bx\cdot F_t,~\bx'\cdot F_t)&=\min\Big\{ |\p|:~\exists \text{ an error vector } \bz\in\p \text{ s.t. } (\bx-\bx')\cdot F_t=\bz\cdot G_t \Big\}\nonumber\\
&=\min\Big\{ |\p|:~ (\bx-\bx')\cdot F_t\in \Delta(t,\p) \Big\}\nonumber\\
&=d^{(t)}\big(\bzero,~(\bx-\bx')\cdot F_t\big),
\end{align}
where $\bzero$ stands for the all-zero row $|\In(t)|$-vector. In the rest of the paper, we always use~$\bzero$ to denote an all-zero (row or column) vector in the paper, whose dimension should be clear from the context. By~\eqref{equ5}, we rewrite $d_{\min}^{(t)}$ as:
\begin{align}
d_{\min}^{(t)}
&=\min_{\bx, \bx'\in \Fq^{\w}:~\atop \bx \neq \bx'} d^{(t)}\big(\bzero,~(\bx-\bx')\cdot F_t\big)\nonumber\\
&=\min_{\bx \in \Fq^{\w}\setminus\{\bzero\}}~ d^{(t)}\big(\bzero,~\bx\cdot F_t\big)\nonumber\\
&=\min_{\bx \in \Fq^{\w}\setminus\{\bzero\}} \min\Big\{ |\p|:~ \bx\cdot F_t\in \Delta(t,\p) \Big\}\nonumber\\
&=\min\Big\{ |\p|:~ \Phi(t)\bigcap \Delta(t,\p)\neq \{\bzero\} \Big\}.\label{equ6}
\end{align}
In the rest of the paper, we use \eqref{equ6} as the definition of the minimum distance of a rate-$\w$ LNEC code $\tC$ on the network $G$ at the sink node $t\in T$, which is more convenient for discussion. We thus write this definition as follows.
\begin{defn}\label{def:min_dist}
Consider a rate-$\w$ LNEC code $\tC$ on the network $G$. The minimum distance of $\tC$ at a sink node $t$ is defined as
\begin{align}\label{equ:d_min_t}
d_{\min}^{(t)}=\min\Big\{ |\p|:~\Phi(t)\bigcap\Delta(t,\p)\neq\{\bzero\} \Big\}.
\end{align}
\end{defn}

Furthermore, it is not difficult to see that the distance $d^{(t)}(\cdot, \cdot)$ defined in~\eqref{distance} is equivalent to the distance measure defined in Definition~1 in \cite{Yang-refined-Singleton}, while the minimum distance $d_{\min}^{(t)}$ defined in~\eqref{equ:d_min_t} is the same as the minimum distance defined in Definition~7 in \cite{zhang-correction} (see Proposition~2 in \cite{Guang-MDS}). Thus, the 2 LNEC approaches developed in \cite{Yang-refined-Singleton} and \cite{zhang-correction} are in fact equivalent.

For a rate-$\w$ LNEC code $\tC$, the minimum distance $d_{\min}^{(t)}$ at each sink node $t\in T$ characterizes its error correction capability. More precisely, $\tC$ can correct up to $\left\lfloor \big(d_{\min}^{(t)}-1\big)/2 \right\rfloor$ errors at each sink node $t\in T$ (cf.~\cite{Yeung-Cai-correct-1,zhang-correction,Yang-refined-Singleton,Guang-MDS,Guang-Zhang-NECBook}).

To see this, we consider 2 arbitrary pairs $(\bx_1~\bz_1)$ and $(\bx_2~\bz_2)$ of source message vector and error vector such that the Hamming weight $w_H(\bz_i)\leq \left\lfloor \big(d_{\min}^{(t)}-1\big)/2 \right\rfloor$, $i=1,2$, and
$(\bx_1~\bz_1)\cdot \F_t=(\bx_2~\bz_2)\cdot \F_t$,
or equivalently,
\begin{align}\label{equ1:prop:er-corrt_capability}
(\bx_1-\bx_2)\cdot F_t=(\bz_2-\bz_1)\cdot G_t.
\end{align}
Let $\p_i=\big\{ e\in E:~z_{i,e} \neq 0 \big\}$, where $\bz_i \triangleq \big(z_{i,e}:~e\in E \big)$, $i=1,2$. Clearly, $\bz_i \in \p_i$ and $|\p_i|\leq  \left\lfloor \big(d_{\min}^{(t)}-1\big)/2 \right\rfloor$, $i=1,2$.
Further, let $\p=\p_1\bigcup\p_2$. Then,
\begin{align*}
|\p|\leq |\p_1|+|\p_2|\leq d_{\min}^{(t)}-1,
\end{align*}
and $\bz_2-\bz_1 \in \p$.
By the definition of $d_{\min}^{(t)}$ (cf.~\eqref{equ:d_min_t}), we immediately have
\begin{align}\label{equ2-prop:er-corrt_capability}
\Phi(t)\bigcap\Delta(t,\p)=\{\bzero\}.
\end{align}
Together with $(\bx_1-\bx_2)\cdot F_t \in \Phi(t)$ and $(\bz_2-\bz_1)\cdot G_t \in \Delta(t,\p)$, we obtain that
\begin{align*}
(\bx_1-\bx_2)\cdot F_t=(\bz_2-\bz_1)\cdot G_t =\bzero.
\end{align*}
It thus follows from $\Rank(F_t)=\w$ that $\bx_1 = \bx_2$. In other words, $\tC$ can correct up to $\left\lfloor \big(d_{\min}^{(t)}-1\big)/2 \right\rfloor$ errors at each $t\in T$. We state this result formally in the next theorem. Let $r$ be a nonnegative integer and $\mH(r)$ be the collection of all edge subsets of size up to $r$, i.e.,
\begin{align}\label{Hr}
\mH(r)=\big\{ \p\subseteq E:~ |\p|\leq r \big\}.
\end{align}

\begin{thm}\label{prop:er-corrt_capability}
Consider an $\Fq$-valued rate-$\w$ LNEC code $\tC$ on the network $G$. Let $t$ be a sink node with $\dim\big( \Phi(t) \big)=\w$. At this sink node $t$, the LNEC code $\tC$ can correct any error vector in the set
\begin{align}
\Big\{ \bz\in \Fq^{|E|}:~\bz \in \p \text{ for some } \p\in \mH\big(\left\lfloor \big(d_{\min}^{(t)}-1\big)/2 \right\rfloor\big) \Big\}.
\end{align}
\end{thm}

Next, we present the {\em Singleton bound} on the minimum distance $d_{\min}^{(t)}$ at the sink node $t\in T$:
\begin{align}\label{SingletonBound_t}
d_{\min}^{(t)}\leq C_t-\w+1
\end{align}
(cf.\cite{zhang-correction,Guang-MDS,Yang-refined-Singleton,Guang-Zhang-NECBook}). If an $\Fq$-valued rate-$\w$ LNEC code $\tC$ not only is decodable but also satisfies the Singleton bound~\eqref{SingletonBound_t} with equality for each sink node $t\in T$, i.e.,
\begin{align}\label{MDS-condition}
\dim\big(\Phi(t)\big)=\w \quad \text{and} \quad  d_{\min}^{(t)}=C_t-\w+1, ~~ \forall~t\in T,
\end{align}
then $\tC$ is called {\em maximum distance separable (MDS) for $T$}. Then, in terms of the error correction capability given in Theorem~\ref{prop:er-corrt_capability}, an $\Fq$-valued rate-$\w$ LNEC MDS code has the maximum error correction capability at each sink node.

\section{Enhanced Characterization of LNEC Capability}\label{sec:enhanced_capability}

We first introduce a number of graph-theoretic concepts that will be used frequently in the sequel. We continue to consider a finite directed acyclic network $G=(V,E)$. The {\em reverse network} $G^\top$ of $G$ is obtained from $G$ by reversing the direction of every edge on $G$. It is evident that a subset of $E$ is a cut separating a node $v$ from a node $u$ on $G$ if and only if this subset of $E$ is a cut separating $u$ from $v$ on $G^\top$. Inspired by this observation, for an edge subset $\p$ and a non-source node $u$, a subset of $E$ is called a {\em cut separating $u$ from $\p$} on $G$ if this edge subset is a cut separating~$\p$ from $u$ on $G^\top$ (cf.~Section~\ref{Sec_II-A}). The {\em capacity} of the cut separating $u$ from $\p$ on $G$ is accordingly defined as the number of edges in the cut. The minimum of the capacities of all cuts separating $u$ from $\p$ on $G$ is called the {\em minimum cut capacity} separating~$u$ from $\p$, denoted by $\mincut(\p,u)$. On the network $G$, a cut separating~$u$ from $\p$ is called a {\em minimum cut} separating $u$ from $\p$ if its capacity achieves the minimum cut capacity $\mincut(\p,u)$.

Further, we say that a minimum cut separating $u$ from $\p$ on $G$ is {\em primary} if it separates $u$ from all the minimum cuts that separate $u$ from $\p$ on $G$. The concept of primary minimum cut was introduced by Guang and Yeung~\cite{GY-SNC-Reduction}, where its existence and uniqueness were proved.
Finally, we say that an edge subset $\p$ is {\em primary for $u$} if $\p$ is the primary minimum cut separating $u$ from $\p$. We now use the following example to illustrate these concepts.

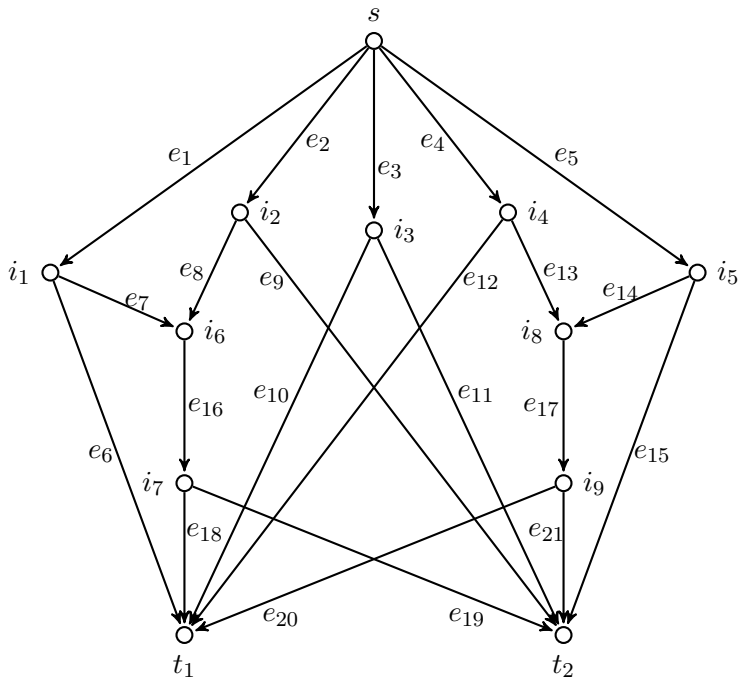
\begin{figure}[!t]
\centering
\begin{tikzpicture}
[->,>=stealth',shorten >=1pt,auto,node distance=2.52cm, thick]
  \tikzstyle{every state}=[fill=none,draw=black,text=black,minimum size=6pt,inner sep=0pt]
  \node[state]         (s)[label=above:$s$]                 {};
  \node[state]         (i_2)[below left of=s, xshift=0mm, yshift=-5mm, label=right:$i_2$]   {};
  \node[state]         (i_4)[below right of=s, xshift=0mm, yshift=-5mm, label=right:$i_4$] {};
  \node[state]         (i_3)[below of=s, yshift=0mm, label=right:$i_3$] {};
  \node[state]         (i_1)[left of=i_2, xshift=0mm, yshift=-8mm, label=left:$i_1$] {};
  \node[state]         (i_5)[right of=i_4, xshift=0mm, yshift=-8mm, label=right:$i_5$] {};
  \node[state]         (i_6)[below right of=i_1, yshift=10mm, label=right:$i_6$] {};
  \node[state]         (i_7)[below of=i_6, yshift=5mm, label=left:$i_7$] {};
  \node[state]         (i_8)[below left of=i_5, yshift=10mm, label=left:$i_8$] {};
  \node[state]         (i_9)[below of=i_8, yshift=5mm, label=right:$i_9$] {};
  \node[state]         (t_1)[below of=i_7, yshift=5mm, label=below:$t_1$] {};
  \node[state]         (t_2)[below of=i_9, yshift=5mm, label=below:$t_2$] {};
\path
(s) edge           node[left=1mm] {$e_1$} (i_1)
    edge           node[pos=0.6, right=0mm] {$e_2$} (i_2)
    edge           node[pos=0.7, right=-1mm] {$e_3$} (i_3)
    edge           node[pos=0.6, left=-0.5mm] {$e_4$} (i_4)
    edge           node[right=1mm] {$e_5$} (i_5)
(i_1) edge         node[left=-1mm]{$e_6$}(t_1)
      edge       node[pos=0.5, right=-0.5mm]{$e_7$}(i_6)
(i_2) edge       node[pos=0.5, left=-0.5mm]{$e_8$}(i_6)
      edge       node[pos=0.15, left=-0.5mm]{$e_9$}(t_2)
(i_3) edge       node[pos=0.4, left=-0.5mm]{$e_{10}$}(t_1)
      edge       node[pos=0.4, right=-0.5mm]{$e_{11}$}(t_2)
(i_4) edge       node[pos=0.15, right=-0.5mm]{$e_{12}$}(t_1)
      edge       node[pos=0.5, right=-0.5mm]{$e_{13}$}(i_8)
(i_5) edge       node[pos=0.3, left=0.5mm]{$e_{14}$}(i_8)
      edge       node[right=-1mm]{$e_{15}$}(t_2)
(i_6) edge       node[pos=0.5, right=-1mm]{$e_{16}$}(i_7)
(i_8) edge       node[pos=0.5, left=-1mm]{$e_{17}$}(i_9)
(i_7) edge       node[pos=0.3, right=-1.2mm]{$e_{18}$}(t_1)
      edge       node[pos=0.9, left=3mm]{$e_{19}$}(t_2)
(i_9) edge       node[pos=0.9, right=3mm]{$e_{20}$}(t_1)
      edge       node[pos=0.3, left=-1.7mm]{$e_{21}$}(t_2);
\end{tikzpicture}
\caption{The network $G$.}
\label{Fig_network}
\end{figure}

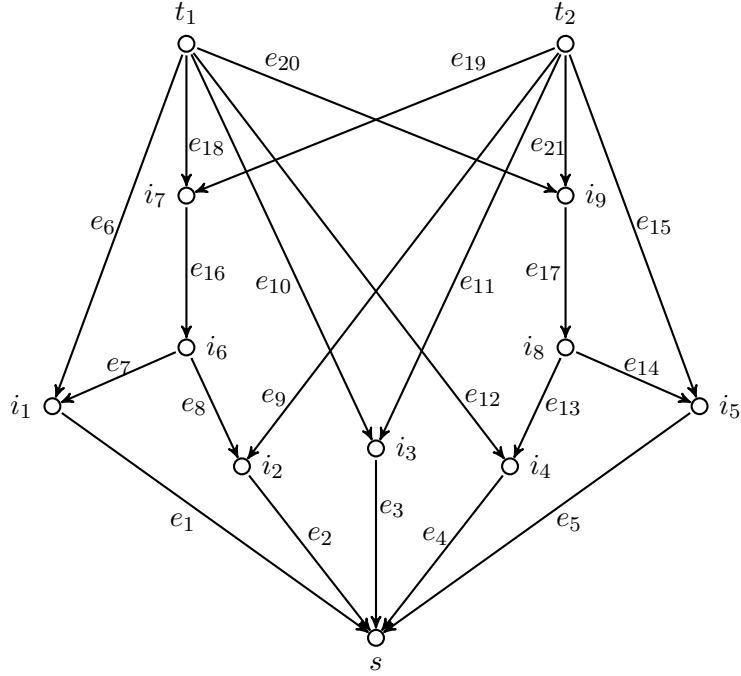
\begin{figure}[!htb]
\centering
\begin{tikzpicture}
[<-,>=stealth',shorten >=1pt,auto,node distance=2.52cm, thick]
  \tikzstyle{every state}=[fill=none,draw=black,text=black,minimum size=6pt,inner sep=0pt]
  \node[state]         (s)[label=below:$s$]                 {};
  \node[state]         (i_2)[above left of=s, xshift=0mm, yshift=5mm, label=right:$i_2$]   {};
  \node[state]         (i_4)[above right of=s, xshift=0mm, yshift=5mm, label=right:$i_4$] {};
  \node[state]         (i_3)[above of=s, yshift=0mm, label=right:$i_3$] {};
  \node[state]         (i_1)[left of=i_2, xshift=0mm, yshift=8mm, label=left:$i_1$] {};
  \node[state]         (i_5)[right of=i_4, xshift=0mm, yshift=8mm, label=right:$i_5$] {};
  \node[state]         (i_6)[above right of=i_1, yshift=-10mm, label=right:$i_6$] {};
  \node[state]         (i_7)[above of=i_6, yshift=-5mm, label=left:$i_7$] {};
  \node[state]         (i_8)[above left of=i_5, yshift=-10mm, label=left:$i_8$] {};
  \node[state]         (i_9)[above of=i_8, yshift=-5mm, label=right:$i_9$] {};
  \node[state]         (t_1)[above of=i_7, yshift=-5mm, label=above:$t_1$] {};
  \node[state]         (t_2)[above of=i_9, yshift=-5mm, label=above:$t_2$] {};
\path
(s) edge           node[left=1mm] {$e_1$} (i_1)
    edge           node[pos=0.6, right=0mm] {$e_2$} (i_2)
    edge           node[pos=0.7, right=-1mm] {$e_3$} (i_3)
    edge           node[pos=0.6, left=-0.5mm] {$e_4$} (i_4)
    edge           node[right=1mm] {$e_5$} (i_5)
(i_1) edge         node[left=-1mm]{$e_6$}(t_1)
      edge       node[pos=0.7, left=0mm]{$e_7$}(i_6)
(i_2) edge       node[pos=0.5, left=-0.5mm]{$e_8$}(i_6)
      edge       node[pos=0.15, left=-0.5mm]{$e_9$}(t_2)
(i_3) edge       node[pos=0.4, left=-0.5mm]{$e_{10}$}(t_1)
      edge       node[pos=0.4, right=-0.5mm]{$e_{11}$}(t_2)
(i_4) edge       node[pos=0.15, right=-0.5mm]{$e_{12}$}(t_1)
      edge       node[pos=0.5, right=-0.5mm]{$e_{13}$}(i_8)
(i_5) edge       node[pos=0.7, right=0.5mm]{$e_{14}$}(i_8)
      edge       node[right=-1mm]{$e_{15}$}(t_2)
(i_6) edge       node[pos=0.5, right=-1mm]{$e_{16}$}(i_7)
(i_8) edge       node[pos=0.5, left=-1mm]{$e_{17}$}(i_9)
(i_7) edge       node[pos=0.3, right=-1.2mm]{$e_{18}$}(t_1)
      edge       node[pos=0.9, left=3mm]{$e_{19}$}(t_2)
(i_9) edge       node[pos=0.9, right=3mm]{$e_{20}$}(t_1)
      edge       node[pos=0.3, left=-1.7mm]{$e_{21}$}(t_2);
\end{tikzpicture}
\caption{The reverse network $G^\top$ of $G$.}
\label{Fig_reverse_network}
\end{figure}

\begin{example}\label{EX_1}

Consider the network $G$ depicted in Fig.~\ref{Fig_network}. On the network $G$, we consider an edge subset $\p=\{e_2,e_5\}$ and a node $t_1$. On the reverse network $G^\top$ of $G$ depicted in Fig.~\ref{Fig_reverse_network}, we note that the edge subset $\eta=\{ e_{14}, e_{16}\}$ is a cut separating $\p$ from $t_1$. So the edge subset $\eta$ is a cut separating $t_1$ from $\p$ on $G$ (see Fig.~\ref{Fig_network}). It can be checked that $\eta$ is actually a minimum cut separating $t_1$ from $\p$ on $G$.
Furthermore, the unique primary minimum cut on $G$ separating $t_1$ from $\p$ is the edge subset $\{ e_{18}, e_{20}\}$, which implies that $\{ e_{18}, e_{20}\}$ is primary for $t_1$.

\end{example}

We now consider a sink node $t$ on the network $G$. Let $r$ be a nonnegative integer not larger than $C_t$, the minimum cut capacity separating $t$ from the source node $s$. We define the following two collections of edge subsets on $G$,
\begin{align}
\msE_t(r)=\Big\{ \p\subseteq E:\ \mincut(\p,t)\leq r \Big\}
\end{align}
and
\begin{align}\label{A_t(r,G)}
\mA_t(r)=\Big\{\p\subseteq E:\ |\p|=r \text{ and $\p$ is primary for $t$} \Big\}.\footnotemark
\end{align}
\footnotetext{Note that $\mA_t(r)=\msE_t(r)=\emptyset$ when $r=0$.}
We now present the following theorem which is one of the main results of this paper.

\begin{thm}\label{lem_equi_Edt_Adt}
Consider a rate-$\w$ LNEC code over a finite field $\mF$ on the network $G$. Then for a sink node $t$ with $C_t\geq \w$ and a nonnegative integer $r\leq C_t-\w$,
\begin{align}\label{equ-lem_equi_Edt}
\Phi(t)\bigcap\Delta(t,\p)=\{\bzero\},\qquad \forall~ \p \in \msE_t(r),
\end{align}
if and only if
\begin{align}\label{equ-lem_equi_Adt}
\Phi(t)\bigcap\Delta(t,\p)=\{\bzero\},\qquad \forall~ \p \in \mA_t(r).
\end{align}
\end{thm}

In order to prove Theorem~\ref{lem_equi_Edt_Adt}, we need the following lemma.

\begin{lemma}\label{lem_for_prop}
For an edge subset $\p$ and a sink node $t$, and any integer $r$ such that $\mincut(\p, t)\leq r \leq C_t$,\footnote{If $C_t < \mincut(\p, t)$, then no such integer $r$ exists.} there exists a size-$r$ primary edge subset $\eta$ for~$t$ such that $\eta$ separates $t$ from $\p$.
\end{lemma}
\begin{IEEEproof}
Consider an arbitrary edge subset $\p$ with $\mincut(\p, t)\leq r$. For the case of $\mincut(\p, t)=r$, the lemma is evidently true by the existence of the primary minimum cut separating $t$ from $\p$. It thus suffices to consider the case of $\mincut(\p, t)<r$. For this case, since $r \leq C_t$, we claim that there exists an edge subset $\widehat{\p}$ satisfying $\p \subseteq \widehat{\p}$ and $\mincut(\widehat{\p}, t)=r$. Indeed, note that when we add an edge $e$ to $\p$, the minimum cut capacity $\mincut(\p\bigcup\{e\}, t)$ separating $t$ from $\p\bigcup\{e\}$ satisfies
\begin{align}\label{equ_Lem4}
\mincut\big(\p, t\big) \leq \mincut\big(\p\bigcup\{e\}, t\big) \leq \mincut\big(\p, t\big)+1,
\end{align}
i.e., the minimum cut capacity can be increased at most by $1$. Note that $\p\subseteq E$ and we have $\mincut\big(E, t\big)\geq C_t$. Thus we see that for any $r \leq C_t$, in view of~\eqref{equ_Lem4}, we can always add edges to $\p$ one by one to form an edge subset $\widehat{\p}$ until $\mincut(\widehat{\p}, t)=r$. Clearly, $\p \subseteq \widehat{\p}$. Thus, the primary minimum cut separating $t$ from $\widehat{\p}$, denoted by $\eta$, separates $t$ from $\p$. Together with the fact that the primary minimum cut $\eta$ separating $t$ from $\widehat{\p}$ is primary for $t$ and $|\eta|=\mincut(\widehat{\p}, t)=r$, the lemma is proved.
\end{IEEEproof}\medskip

With Lemma~\ref{lem_for_prop}, we are now ready to prove Theorem~\ref{lem_equi_Edt_Adt}.

\begin{IEEEproof}[Proof of Theorem~\ref{lem_equi_Edt_Adt}]
The ``only if'' part (i.e., \eqref{equ-lem_equi_Edt} $\Rightarrow$ \eqref{equ-lem_equi_Adt}) is evident since $\mA_t(r) \subseteq \msE_t(r)$. We now prove the ``if'' part (i.e., \eqref{equ-lem_equi_Adt} $\Rightarrow$ \eqref{equ-lem_equi_Edt}). We consider an arbitrary edge subset $\p \in \msE_t(r)$. Then,
\begin{align*}
\mincut\big(\p, t\big)\leq r\leq C_t-\w \leq C_t.
\end{align*}
By Lemma~\ref{lem_for_prop}, there exists a primary edge subset $\eta$ in $\mA_t(r)$ such that $\eta$ separates $t$ from $\p$. For a directed path $P=(e_1,e_2,\cdots,e_m)$, $m\geq 1$, on the extended network $\wtG$, we define
\begin{align}\label{K_P}
K_P=
 \begin{cases}
    1 & \text{ if $m=1$}; \\
    \prod\limits_{i=1}^{m-1} k_{e_i,e_{i+1}}  & \text{ if $m\geq 2$.}
  \end{cases}
\end{align}

We consider an imaginary error edge $e'\in E'$, which is associated with the edge $e\in E$, and an edge $\e\in E$. By calculating by~\eqref{equ_ext_f} recursively according to the given ancestral order on the edges in $E$, it is not difficult to obtain that
\begin{align}\label{pf_Them3_equ1}
\f_{\e}(e')=\sum_{P:~\text{a directed path from $e'$ to $\e$}}K_P,
\end{align}
where if no directed paths exist from $e'$ to $\e$, we can see that $\f_{\e}(e')=0$ by~\eqref{pf_Them3_equ1}.
Continuing from~\eqref{pf_Them3_equ1}, we obtain that
\begin{align}
\f_{\e}(e')=&\sum_{P:~\text{a directed path from $e'$ to $\e$} \atop \text{\quad passing through the edge $e$}} K_{P}+\sum_{P:~\text{a directed path from $e'$ to $\e$} \atop \text{\qquad not passing through the edge $e$}} K_{P}\label{pf_Them3_equ2}\\
=&\sum_{P:~\text{a directed path from $e'$ to $\e$} \atop \text{\quad passing through the edge $e$}} K_{P},\label{L_e}
\end{align}
where the last equality~\eqref{L_e} is justified as follows. First, if no directed paths exist from $e'$ to $\e$, then we easily see that
\begin{align*}
\sum_{P:~\text{a directed path from $e'$ to $\e$} \atop \text{\quad passing through the edge $e$}} K_{P}=0=\f_{\e}(e').
\end{align*}
Thus, the equality~\eqref{L_e} is satisfied. Otherwise, we consider the two cases below.

\noindent\textbf{Case 1:} $e=\e$.

In this case, we note that $(e',e)$ is the unique directed path from $e'$ to $e$ by the acyclicity of the extended network $\widetilde{G}$. Then, there does not exist a path from $e'$ to $e$ not passing through $e$. This immediately implies that the second term in $\eqref{pf_Them3_equ2}$ is $0$, i.e.,
\begin{align*}
\sum_{P:~\text{a directed path from $e'$ to $\e$} \atop \text{\qquad not passing through the edge $e$}} K_{P}=0.
\end{align*}
We thus have proved the equality~\eqref{L_e} in this case. Further, we have
\begin{align*}
\f_{e}(e')=\sum_{P:~\text{a directed path from $e'$ to $e$} \atop \text{\quad passing through the edge $e$}} K_{P}=K_{(e',e)}=k_{e',e}=1.
\end{align*}

\noindent\textbf{Case 2:} $e \neq \e$.

If there does not exist a path from $e'$ to $\e$ not passing through $e$, similar to the above discussion in Case~1, the second term in $\eqref{pf_Them3_equ2}$ is $0$ and so we have proved the equality~\eqref{L_e}. Otherwise, each directed path $P$ from $e'$ to $\e$ not passing through the edge $e$ can be regarded as the concatenation of two sub-paths, where one is a length-$2$ path $(e',c)$ from $e'$ to some edge $c\in \Out(\tail(e))\setminus\{e\}$; the other is a directed path from $c$ to $\e$, denoted by $P_{c \rightarrow \e}$. Note that these two paths overlap on the edge $c$. Together with $k_{e',c}=0$ as $c \in \Out(\tail(e))\setminus\{e\}$ (cf.~\eqref{k_e'_d}), it follows from~\eqref{K_P} that
\begin{align}\label{pf_Them3_equ3}
K_P=K_{(e',c)} \cdot K_{P_{c \rightarrow \e}}=k_{e',c }\cdot K_{P_{c \rightarrow \e}}=0\cdot K_{P_{c \rightarrow \e}}=0.
\end{align}
This implies that the second term in $\eqref{pf_Them3_equ2}$ is $0$ and thus we have proved the equality~\eqref{L_e}. In particular, we note that the above argument also applies to the special case that $\e \in \Out(\tail(e))\setminus\{e\}$. To be specific, in~\eqref{pf_Them3_equ3}, $K_{P_{c \rightarrow \e}}=1$ if $c=\e$ (cf.~\eqref{K_P}).


\bigskip

Now, continuing from~\eqref{L_e}, we have
\begin{align}
\f_{\e}(e')&=\sum_{P:~\text{a directed path from $e'$ to $\e$ } \atop \text{\quad passing through the edge $e$}} K_{P}\nonumber\\
&=\sum_{P:~\text{a directed path from $e$ to $\e$}} k_{e',e} \cdot K_{P}\nonumber\\
&=\sum_{P:~\text{a directed path from $e$ to $\e$}} K_{P},\label{equ7}
\end{align}
where~\eqref{equ7} also follows from $k_{e',e}=1$ (cf.~\eqref{k_e'_d}). Note that \eqref{equ7} continues to hold when there exists no directed path from $e'$ to $\e$.

Next, we will prove that $\Delta(t,\p)\subseteq\Delta(t,\eta)$, where we recall that $\p$ is any edge subset in $\msE_t(r)$ and $\eta$ is any primary edge subset in $\mA_t(r)$ such that $\eta$ separates $t$ from $\p$. Toward this end, we consider two cases for an edge $e\in \p$.

\noindent\textbf{Case 1:} $e \in \In(t)$, i.e., $e\in \p\bigcap \In(t)$.

We first claim that $e\in \eta$, because otherwise $\eta$ cannot separate $t$ from $\{e\}$ (which is a subset of $\p$) and thus cannot separate $t$ from $\p$, a contradiction.
Now, we consider the row vector $\row_t(e')=\big( \f_{\e}(e'):~ \e \in \In(t) \big)$, where $e'$ is the imaginary error edge associated with $e$. By the above claim that $e\in \eta$, we immediately prove that $\row_t(e')\in \Delta(t,\eta)$ (cf.~\eqref{equ-source_error_spaces}).

\noindent\textbf{Case 2:} $e \notin \In(t)$, i.e., $e\in \p\setminus  \In(t)$.

We consider an arbitrary edge $\e\in \In(t)$. If there exists a directed path $P$ from $e$ to $\e$, then this path $P$ has length at least $2$ and can be regarded as the concatenation of two sub-paths, where one is a length-$2$ path $(e,d)$ from $e$ to some edge $d\in \Out(\head(e))$; the other is a directed path $P_{d\rightarrow \e}$ from $d$ to $\e$. By~\eqref{K_P}, we have
\begin{align}\label{K_P-2}
K_P=K_{(e,d)} \cdot K_{P_{d\rightarrow \e}}= k_{e,d}\cdot K_{P_{d\rightarrow \e}}.
\end{align}
On the other hand, if there exists no path from $e$ to $\e$, then we readily see that for any edge $d\in \Out(\head(e))$, there exists no path from $d$ to $\e$, either.

Then, continuing from~\eqref{equ7}, we obtain that
\begin{align}
\f_{\e}(e')=&\sum_{d\in \Out(\head(e))}~\sum_{P:~\text{a directed path }\atop \text{\quad from $e$ to $\e$ via $d$}} K_{P}\nonumber\\
=&\sum_{\text{$d\in \Out(\head(e))$ :}\atop \text{$\exists$ a path from $e$ to $\e$ via $d$}}~\sum_{P:~\text{a directed path }\atop \text{\quad from $e$ to $\e$ via $d$}} K_{P}+
\sum_{\text{$d\in \Out(\head(e))$ :}\atop \text{$\nexists$ a path from $e$ to $\e$ via $d$}}~\sum_{P:~\text{a directed path }\atop \text{\quad from $e$ to $\e$ via $d$}} K_{P} \label{equ1-lem_equi_Edt_Adt-case2}\\
=&\sum_{\text{$d\in \Out(\head(e))$ :}\atop \text{$\exists$ a path from $e$ to $\e$ via $d$}}
k_{e,d}\cdot \bigg(\sum_{P_{d\rightarrow \e}:~\text{a directed path from $d$ to $\e$}} K_{P_{d\rightarrow \e}} \bigg)\nonumber \\
& +\sum_{\text{$d\in \Out(\head(e))$ :}\atop \text{$\nexists$ a path from $e$ to $\e$ via $d$}}k_{e,d}\cdot \bigg(\sum_{P_{d\rightarrow \e}:~\text{a directed path from $d$ to $\e$}} K_{P_{d\rightarrow \e}} \bigg),\label{equ2-lem_equi_Edt_Adt-case2}
\end{align}
where the last equality~\eqref{equ2-lem_equi_Edt_Adt-case2} is explained as follows. We first consider the first term in~\eqref{equ1-lem_equi_Edt_Adt-case2}. By \eqref{K_P-2}, we immediately obtain that
\begin{align}\label{equ3-lem_equi_Edt_Adt-case2}
\sum_{\text{$d\in \Out(\head(e))$ :}\atop \text{$\exists$ a path from $e$ to $\e$ via $d$}}~\sum_{P:~\text{a directed path }\atop \text{\quad from $e$ to $\e$ via $d$}} K_{P}
=\sum_{\text{$d\in \Out(\head(e))$ :}\atop \text{$\exists$ a path from $e$ to $\e$ via $d$}}
k_{e,d}\cdot \bigg(\sum_{P_{d\rightarrow \e}:~\text{a directed path from $d$ to $\e$}} K_{P_{d\rightarrow \e}} \bigg).
\end{align}
Next, we consider the second term in~\eqref{equ1-lem_equi_Edt_Adt-case2}. We note that for an edge $d\in \Out(\head(e))$, there exists no path from $e$ to $\e$ via $d$ if and only if there exists no path from $d$ to $\e$. As such, we obtain that
\begin{align}\label{equ4-lem_equi_Edt_Adt-case2}
\sum_{\text{$d\in \Out(\head(e))$ :}\atop \text{$\nexists$ a path from $e$ to $\e$ via $d$}}~\sum_{P:~\text{a directed path }\atop \text{\quad from $e$ to $\e$ via $d$}} K_{P}=0
\end{align}
and
\begin{align}\label{equ5-lem_equi_Edt_Adt-case2}
\sum_{\text{$d\in \Out(\head(e))$ :}\atop \text{$\nexists$ a path from $e$ to $\e$ via $d$}}k_{e,d}\cdot \bigg(\sum_{P_{d\rightarrow \e}:~\text{a directed path from $d$ to $\e$}} K_{P_{d\rightarrow \e}} \bigg)=0.
\end{align}
Combining \eqref{equ3-lem_equi_Edt_Adt-case2}, \eqref{equ4-lem_equi_Edt_Adt-case2} and \eqref{equ5-lem_equi_Edt_Adt-case2}, we immediately prove the equality~\eqref{equ2-lem_equi_Edt_Adt-case2}, and we further obtain that
\begin{align}\label{equ9}
\f_{\e}(e')&=\sum_{d\in \Out(\head(e))} k_{e,d}\cdot \bigg(\sum_{P:~\text{a directed path from $d$ to $\e$}} K_{P} \bigg)\nonumber\\
&=\sum_{d\in \Out(\head(e))} k_{e,d}\cdot \f_{\e}(d'),
\end{align}
where the equality~\eqref{equ9} again follows from~\eqref{equ7} with $d$ in place of $e$. In particular, the equality~\eqref{equ9} holds when there exists no path from $e'$ to $\e$, with
\begin{align*}
\f_{\e}(e')=0 \quad \text{ and }\quad  \f_{\e}(d')=0,~\forall~d\in \Out(\head(e)).
\end{align*}

Now, for the row vector $\row_t(e')=\big( \f_{\e}(e'):~ \e \in \In(t) \big)$, by~\eqref{equ9} we obtain that
\begin{align}\label{equ_equality_row}
\row_t(e')&=\big( \f_{\e}(e'):~ \e \in \In(t) \big)\nonumber\\
         &=\bigg( \sum_{d\in \Out(\head(e))} k_{e,d}\cdot \f_{\e}(d'):~ \e \in \In(t) \bigg)\nonumber\\
         &=\sum_{d\in \Out(\head(e))} k_{e,d}\cdot \Big( \f_{\e}(d'):~ \e \in \In(t) \Big)\nonumber\\
         &=\sum_{d\in \Out(\head(e))} k_{e,d}\cdot \row_t(d').
\end{align}
Further, for any $d\in \Out(\head(e))$, if no path exists from $d$ to the sink node $t$, by~\eqref{equ7} we have
\begin{align*}
\f_{\e}(d')=0,~\forall~\e \in \In(t),
\end{align*}
implying that $\row_t(d')=\bzero$. Thus, continuing from~\eqref{equ_equality_row}, we obtain that
\begin{align}\label{equ1_equality_row}
\row_t(e')&=\sum_{d\in \Out(\head(e))} k_{e,d}\cdot \row_t(d')\nonumber\\
&=\sum_{d\in \Out(\head(e)) \text{:}\atop \text{$\exists$ a path from $d$ to $t$}} k_{e,d}\cdot \row_t(d').
\end{align}

In~\eqref{equ1_equality_row}, for each $d$ in the summation, apply~\eqref{equ1_equality_row} recursively for $\row_t(d')$ by letting $e$ be $d$ until all the edges $d$ in the summation are in $\eta$.  Then we obtain that $\row_t(e')$ is a linear combination of $\row_t(d')$, $d\in \eta$, i.e., $\row_t(e')\in \Delta(t,\eta)$.

\bigskip

Now, we combine the above two cases and immediately obtain that $\row_t(e')\in \Delta(t,\eta)$ for all $e\in \p$, or equivalently, $\Delta(t,\p)\subseteq\Delta(t,\eta)$. Then \eqref{equ-lem_equi_Adt} implies that $\Phi(t)\bigcap\Delta(t,\p)=\{\bzero\}$. We thus have proved the ``if'' part and also the theorem.
\end{IEEEproof}\medskip

Recall the definition of $\mH(r)$ in~\eqref{Hr}. We immediately obtain the following corollary.

\begin{cor}\label{cor-10}
Consider a rate-$\w$ LNEC code over a finite field $\mF$ on the network $G$. Then for a sink node~$t$ with $C_t\geq \w$ and a nonnegative integer $r\leq C_t-\w$, the conditions~\eqref{equ-lem_equi_Edt}, \eqref{equ-lem_equi_Adt} and the condition
\begin{align}\label{equ-lem_equi_Hdt}
\Phi(t)\bigcap\Delta(t,\p)=\{\bzero\},\qquad \forall~ \p \in \mH(r)
\end{align}
are all equivalent.
\end{cor}
\begin{IEEEproof}
Note that
\begin{align*}
\mA_t(r) \subseteq \mH(r) \subseteq \msE_t(r).
\end{align*}
Hence, we obtain that \eqref{equ-lem_equi_Edt} $\Rightarrow$ \eqref{equ-lem_equi_Hdt} and \eqref{equ-lem_equi_Hdt} $\Rightarrow$ \eqref{equ-lem_equi_Adt}. Together with \eqref{equ-lem_equi_Adt}  $\Leftrightarrow$ \eqref{equ-lem_equi_Edt} from Theorem~\ref{lem_equi_Edt_Adt}, the corollary is proved.
\end{IEEEproof}\medskip

Together with the equivalence of \eqref{equ-lem_equi_Edt} and \eqref{equ-lem_equi_Adt} in Theorem~\ref{lem_equi_Edt_Adt} and the discussion above Definition~\ref{defn_distance}, we see that at a sink node~$t$, the ``effect'' of any error vector $\bz\in \p$ for an edge subset $\p \in \msE_t(r)$ is equal to the ``effect'' of an error vector $\bz'\in \eta$ for a primary edge subset $\eta \in \mA_t(r)$ such that $\eta$ separates $t$ from $\p$, i.e., $\bz\cdot G_t=\bz'\cdot G_t$. Thus, to ensure that an LNEC code $\tC$ can correct any error vector in the set of error vectors
\begin{align}\label{Z_msE_t_r}
\mZ\big(\msE_t(r)\big)\triangleq \Big\{ \bz\in \Fq^{|E|}:~\bz \in \p \text{ for some } \p \in \msE_t(r) \Big\},
\end{align}
we only need to ensure that the code $\tC$ can correct any error vector in the reduced set of error vectors
\begin{align}\label{Z_mA_t_r}
\mZ\big(\mA_t(r)\big)\triangleq\Big\{ \bz\in \Fq^{|E|}:~ \bz \in \p \text{ for some } \p\in \mA_t(r) \Big\}.
\end{align}
Thus we have proved the following important consequence.



\begin{thm}\label{thm_A_t}
Consider an $\Fq$-valued rate-$\w$ LNEC code on a network $G=(V,E)$. For a sink node $t\in T$ with $\dim\big( \Phi(t) \big)=\w$, the LNEC code can correct at $t$ any error vector in the set $\mZ\big(\mA_t(r)\big)$ if and only if this code can correct at $t$ any error vector in the set $\mZ\big(\msE_t(r)\big)$.
\end{thm}

By combining Theorem~\ref{lem_equi_Edt_Adt} with Theorem~\ref{thm_A_t}, we immediately enhances Theorem~\ref{prop:er-corrt_capability} in the following corollary.

\begin{cor}\label{thm_enhance_error-correcting-capability}
Consider an $\Fq$-valued rate-$\w$ LNEC code on the network $G$. For a sink node $t\in T$ with $\dim\big( \Phi(t) \big)=\w$, the LNEC code can correct any error vector in the following set of error vectors
\begin{align}
\mZ\Big(\msE_t\big(\left\lfloor \big(d_{\min}^{(t)}-1\big)/2 \right\rfloor\big)\Big)=\Big\{ \bz\in \Fq^{|E|}:~\bz \in \p \text{ for some } \p\in \msE_t\big(\left\lfloor \big(d_{\min}^{(t)}-1\big)/2 \right\rfloor\big) \Big\}.
\end{align}
\end{cor}
\begin{IEEEproof}
By Theorem~\ref{prop:er-corrt_capability}, a rate-$\w$ LNEC code $\tC$ can correct at the sink node $t$ any error vector in the set
\begin{align*}
\Big\{ \bz\in \Fq^{|E|}:~\bz \in \p \text{ for some } \p\in \mH\big(r^*\big) \Big\},
\end{align*}
where we let $r^*=\left\lfloor \big(d_{\min}^{(t)}-1\big)/2\right\rfloor$ for notational simplicity. It follows from $\mA_t(r^*) \subseteq \mH(r^*)$ that the LNEC code $\tC$ can correct at $t$ any error vector in the set $\mZ\big(\mA_t(r^*)\big)$.
By Theorem~\ref{thm_A_t}, $\tC$ can correct at $t$ any error vector in the set $\mZ\big(\msE_t(r^*)\big)$. We thus have proved the corollary.
\end{IEEEproof}\medskip

We now use the following example to illustrate the enhanced characterization of the capability of an LNEC code as asserted in Theorem~\ref{thm_A_t} and Corollary~\ref{thm_enhance_error-correcting-capability}.

\begin{example}
Recall the network $G=(V,E)$ depicted in Fig.~\ref{Fig_network}, where $s$ is the single source node and  $T=\{t_1, t_2\}$ is the set of sink nodes with $C_{t_1}=C_{t_2}=5$. We consider a rate-$3$ LNEC code $\tC$ on $G$ such that $d_{\min}^{(t_1)}=d_{\min}^{(t_2)}=3$. Such a code exists because it satisfies the Singleton bound in \eqref{SingletonBound_t}.

Due to the symmetry of the problem, we only consider the sink node $t_1$ and let $r=\left\lfloor \big(d_{\min}^{(t_1)}-1\big)/2\right\rfloor=1$. We say an edge subset $\p\subseteq E$ is {\em $t_1$-correctable} for this LNEC code $\tC$ if any error vector $\bz\in \p$ can be corrected at $t_1$ in using $\tC$. It follows from Theorem~\ref{prop:er-corrt_capability} that all $21$ edge subsets in
$\mH(1)=\big\{\p\subseteq E:~|\p|\leq 1 \big\}$ are $t_1$-correctable, where clearly, $|\mH(1)|=|E|=21$.

We now consider the enhanced characterization of the capability of an LNEC code $\tC$ in terms of $\msE_{t_1}(r)$ and $\mA_{t_1}(r)$ (cf.~Theorem~\ref{thm_A_t} and Corollary~\ref{thm_enhance_error-correcting-capability}). We first partition $E$ into two edge-disjoint sets
\begin{align*}
E_{t_1}^c\triangleq \big\{  e_{9}, e_{11}, e_{15}, e_{19}, e_{21} \big\}\quad \text{ and }\quad E\setminus E_{t_1}^c.
\end{align*}
Note that $E_{t_1}^c$ is precisely the set of edges in $E$ such that there exists no path from this edge to~$t_1$. Accordingly, $\mH(1)$ is partitioned into two disjoint collections of size-$1$ edge subsets
\begin{align*}
\big\{  \{e_{9}\}, \{e_{11}\}, \{e_{15}\}, \{e_{19}\}, \{e_{21}\} \big\}\quad \text{ and }\quad
\big\{ \{e\}:~ e\in E\setminus E_{t_1}^c \big\}.
\end{align*}

The set of all size-$1$ primary edge subsets for~$t_1$ is given by
\begin{align*}
\mA_{t_1}(1)=\Big\{
      \{ e_{1} \}, \{ e_{4} \}, \{ e_{6} \}, \{ e_{10} \}, \{ e_{12} \}, \{ e_{18} \}, \{ e_{20} \}  \Big\}.
\end{align*}
Consider all the $16$ size-$1$ edge subsets, each of which consists of one edge in $E\setminus E_{t_1}^c$. We see that $\{ e_{10} \}$ is the primary minimum cut separating $t_1$ from $\{e_{3} \}$; $\{ e_{18} \}$ is the primary minimum cut separating $t_1$ from $\{e_{2} \}$, $\{e_{7} \}$, $\{e_{8} \}$ and $\{e_{16}\}$, respectively; and $\{ e_{20} \}$ is the primary minimum cut separating $t_1$ from $\{e_{5} \}$, $\{e_{13} \}$, $\{e_{14} \}$ and $\{e_{17}\}$, respectively. For $i=1,4,6,12$, $\{ e_{i} \}$ is the primary minimum cut separating $t_1$ from only $\{e_{i} \}$ itself.

We write $\p \overset{t_1}{\sim} \eta$ for two edge subsets $\p$ and $\eta$ of $E$ if $\p$ and $\eta$ have the same primary minimum cut with respect to $t_1$, e.g., $\{e_{2} \}\overset{t_1}{\sim}  \{e_{7} \}$, where $\{e_{18}\}$ is the common primary minimum cut separating $t_1$ from $\{e_{2}\}$ and $\{e_{7} \}$. It was proved in \cite{Guang-SmlFieldSize-SNC-comm-lett} that ``\,$\overset{t_1}{\sim}$\,'' is an equivalence relation. With the relation  ``\,$\overset{t_1}{\sim}$\,'', $\mH(1)$ can be partitioned into $8$ equivalence classes
\begin{align}\label{eq-cls}
\begin{split}
&\big\{\{e_{1} \}\big\},\quad \big\{\{e_{4} \}\big\},\quad  \big\{\{e_{6} \}\big\}, \quad \big\{\{e_{12}\}\big\}, \quad \big\{\{e_{3} \}, \{e_{10}\}\big\}, \quad  \big\{ \{e_{2} \}, \{e_{7} \}, \{e_{8} \}, \{e_{16}\}, \{e_{18}\} \big\},\\
&\qquad\quad  \big\{ \{e_{5} \}, \{e_{13} \}, \{e_{14} \}, \{e_{17}\}, \{ e_{20} \} \big\} \quad \text{ and } \quad \big\{ \{e_{9} \}, \{e_{11} \}, \{e_{15} \}, \{e_{19}\}, \{ e_{21} \} \big\},
\end{split}
\end{align}
where for the $5$ edge subsets $\{e_{9}\}$, $\{e_{11} \}$, $\{e_{15} \}$, $\{e_{19}\}$ and $\{e_{21}\}$ in the last equivalence class, the empty set of edges is their common primary minimum cut with respect to $t_1$.

Furthermore, it is not difficult to see that any union of the edge subsets in an equivalence class still have the common primary minimum cut with respect to $t_1$, e.g., $\{e_{18} \}$ is the common primary minimum cut separating $t_1$ from $\{e_{2}, e_{7}\}$ and $\{e_{7}, e_{8}, e_{16}, e_{18}\}$. Moreover, 
for any union of the edge subsets in an equivalence class, say $\p$, and any edge subset $\mu$ of $E_{t_1}^c$ (which is also a union of the edge subsets in the last equivalence class in~\eqref{eq-cls}), we have
\begin{align*}
\mincut(\p \cup \mu,t_1)=\mincut(\p, t_1).
\end{align*}
For example, let $\p=\{e_{2}, e_{7}\}$ and $\mu=\{e_{9}\}$. Then, $\{e_{18}\}$ is the (primary) minimum cut separating $t_1$ from $\p \bigcup \mu$, and
\begin{align*}
\mincut(\p \cup \mu, t_1)=\mincut(\{e_{2}, e_{7}, e_{9}\}, t_1)=\mincut(\{e_{2}, e_{7}\}, t_1)=1.
\end{align*}

Based on the above discussion, by means of a simple calculation, we can obtain that the size of $\mE_{t_1}(1)$ is equal to $2,\!239$,
which is considerably larger than $|\mH(1)|=21$. It follows from Corollary~\ref{thm_enhance_error-correcting-capability} that all the $2,\!239$ nonempty edge subsets in $\mE_{t_1}(1)$ are $t_1$-correctable. On the other hand, by Theorem~\ref{thm_A_t}, in order to ensure that all the $2,\!239$ nonempty edge subsets in $\mE_{t_1}(1)$ are $t_1$-correctable, it suffices to guarantee that the $7$ edge subsets in $\mA_{t_1}(1)$ are $t_1$-correctable.
\end{example}


\section{Field Size Reduction for LNEC Codes}\label{sec:field_size_reduction}


\subsection{Improved Upper Bound on the Minimum Required Field Size}

The minimum required field size for the existence of LNEC codes, particularly LNEC MDS codes, is an open problem not only of theoretical interest but also of practical importance, because it is closely related to the implementation of code constructions in terms of computational complexity and storage requirement. In this subsection, we will present an improved upper bound on the minimum required field size, which shows that the required field size for the existence of LNEC codes in general can be reduced significantly. This new bound is graph-theoretic, which depends only on the network topology and requirement of error correction capability but not on the specific code construction.

\begin{thm}\label{thm_improved_field_size_LNEC_codes}
Let $\mF$ be a finite field of order $q$, where $q$ is a prime power. Let $T$ be the set of sink nodes on the network $G$ with $C_t\geq \w$, $\forall~t\in T$. For each $t\in T$, let $\beta_t$ be a nonnegative integer not larger than $C_t-\w$. Then, there exists an $\Fq$-valued rate-$\w$ LNEC code on $G$ with the minimum distance at $t$ not smaller than $\beta_t+1$ for each $t\in T$, i.e., $d_{\min}^{(t)}\geq \beta_t+1$, $\forall~t\in T$, if the field size $q$ satisfies
\begin{align}\label{equ-field_size_LNEC}
q > \sum_{t \in T}\big|\mA_t(\beta_t)\big|.
\end{align}
\end{thm}

\begin{IEEEproof}
To prove Theorem~\ref{thm_improved_field_size_LNEC_codes}, we need to prove that if \eqref{equ-field_size_LNEC} is satisfied for the field $\mF$, then there exists an $\mF$-valued rate-$\w$ LNEC code on $G$ such that for each sink node $t\in T$,
\begin{align}\label{equ53}
\dim\big(\Phi(t)\big)=\w \quad \text{ and }\quad  d_{\min}^{(t)}\geq \beta_t+1.
\end{align}
By Definition~\ref{def:min_dist}, \eqref{equ53} is equivalent to the condition that for each $t\in T$,
\begin{align}\label{equ54}
\dim\big(\Phi(t)\big)=\w \quad \text{ and }\quad \Phi(t)\bigcap\Delta(t,\p)=\{\bzero\},~~\forall~\p\subseteq E \text{ with } |\p|\leq \beta_t.
\end{align}
We further write the second condition in \eqref{equ54} as
\begin{align*}
\Phi(t)\bigcap\Delta(t,\p)=\{\bzero\}, \quad \forall~\p\in \mH(\beta_t),
\end{align*}
which, by Corollary~\ref{cor-10}, is equivalent to
\begin{align*}
\Phi(t)\bigcap\Delta(t,\p)=\{\bzero\},~~\forall~\p\in \mA_t(\beta_t).
\end{align*}

Based on the above discussion, in order to prove the theorem, it suffices to prove that if the field $\mF$ satisfies \eqref{equ-field_size_LNEC}, then there exists an $\mF$-valued rate-$\w$ LNEC code on $G$ such that for each sink node $t\in T$,
\begin{align}\label{equ-pf-thm_improved_field_size_LNEC_codes}
\dim\big(\Phi(t)\big)=\w \quad \text{ and }\quad \Phi(t)\bigcap\Delta(t,\p)=\{\bzero\},~~\forall~\p \in \mA_t(\beta_t).
\end{align}
This statement can be proved by using a standard argument (e.g.,~the proof of Theorem~1 in \cite{zhang-correction} and the proof of Theorem~5 in \cite{Guang-MDS}). We omit the details here.
\end{IEEEproof}
\smallskip


A straightforward upper bound on the minimum required field size for the existence of a rate-$\w$ LNEC code
with the minimum distance $d_{\min}^{(t)}\geq \beta_t+1$ for each $t\in T$ (where $\beta_t$ is a nonnegative integer not larger than $C_t-\w$) is $\sum_{t \in T} {|E| \choose \beta_t}$.  Such a code can correct at $t$ an arbitrary error vector in the set $\mZ\big(\msE_t(\lfloor \beta_t/2 \rfloor)\big)$ for each $t\in T$. Subsequently, this upper bound was improved in \cite{Guang-MDS} (cf.~\cite[Theorem~8]{Guang-MDS}), as presented in the following proposition. To our knowledge, this is the best known upper bound on the minimum required field size for the existence of such a rate-$\w$ LNEC code.

\begin{prop}\label{original-bound_field_size_LNEC_codes}
Let $\mF$ be a finite field of order $q$, where $q$ is a prime power. Let $T$ be the set of sink nodes on the network $G$ with $C_t\geq \w$, $\forall~t\in T$. For each $t\in T$, let $\beta_t$ be a nonnegative integer not larger than $C_t-\w$. Then, there exists an $\Fq$-valued rate-$\w$ LNEC code on $G$ with the minimum distance $d_{\min}^{(t)}\geq \beta_t+1$ for each $t\in T$ if the field size $q$ satisfies
\begin{align}\label{Fsize-LNEC-GuangFZ}
q > \sum_{t\in T} \big|R_t(\beta_t)\big|,
\end{align}
where
\begin{align}\label{R_t}
R_t(\beta_t)=\big\{ \p\subseteq E:\ |\p|=\mincut(\p,t)=\beta_t \big\}.
\end{align}
\end{prop}

\medskip

We readily see that $\mA_t(\beta_t) \subseteq R_t(\beta_t) \subseteq \big\{ \p\subseteq E:~|\p|=\beta_t \big\}$ and so
\begin{align*}
\sum_{t \in T} \big| \mA_t(\beta_t)\big| \leq \sum_{t \in T} \big| R_t(\beta_t) \big| \leq \sum_{t \in T} {|E| \choose \beta_t}.
\end{align*}
The improvement of our improved bound $\sum_{t \in T} \big| \mA_t(\beta_t)\big|$ in Theorem~\ref{thm_improved_field_size_LNEC_codes} over $\sum_{t \in T} \big| R_t(\beta_t) \big|$ (also over $\sum_{t \in T} {|E| \choose \beta_t}$) is in general significant as illustrated by Example~\ref{eg_field-size} below. The only case when $\sum_{t \in T} \big| \mA_t(\beta_t)\big|$ has no improvement over $\sum_{t \in T} \big| R_t(\beta_t) \big|$, i.e., $\sum_{t \in T} \big| \mA_t(\beta_t)\big| = \sum_{t \in T} \big| R_t(\beta_t) \big|$, is that for each sink node $t$, every edge subset $\p$ with $|\p|=\mincut(\p,t)=\beta_t$ is primary for $t$, i.e., $\p$ is the unique minimum cut separating $t$ from itself. This condition holds only for very special networks. For example, we consider a network as depicted in Fig.~\ref{Fig_classical_coding_model}, where the network consists of only two nodes, a source node $s$ and a sink node $t$, connected by multiple parallel edges from $s$ to $t$. In this network, for any positive integer~$\beta_t$ not larger than $|E|$, i.e., $\beta_t \leq |E|$ (where in fact $|E|$ is the number of multiple parallel edges from $s$ to $t$), we readily see that each edge subset $\p\subseteq E$ of size $\beta_t$ is primary for $t$. This immediately implies that
\begin{align*}
\big| \mA_t(\beta_t)\big| = \big| R_t(\beta_t) \big| =  {|E| \choose \beta_t},\quad \forall~\beta_t\leq |E|.
\end{align*}

\begin{example}\label{eg_field-size}
Recall the network $G=(V,E)$ depicted in Fig.~\ref{Fig_network}, where $s$ is the single source node and $T=\{t_1, t_2\}$ is the set of sink nodes with $C_{t_1}=C_{t_2}=5$. Let the rate $\w=3$ and $\beta_{t_1}=\beta_{t_2}=2$, two nonnegative integers not larger than $C_{t_1}-\w$ and $C_{t_2}-\w$, respectively. We consider an $\Fq$-valued rate-$3$ LNEC code with $d_{\min}^{(t_1)}\geq \beta_{t_1}+1$ and $d_{\min}^{(t_2)}\geq \beta_{t_2}+1$. This code can correct at the sink node $t_i$ an arbitrary error vector in the set $\mZ\big(\msE_{t_i}(1)\big)$ for $i=1,2$. We now focus on the field size $q$ for the existence of such a code.

We first calculate the straightforward bound $\sum_{t \in T} {|E| \choose \beta_t}$ on the field size $q$ as follows:
\begin{align}\label{straightforward-Bound-EX}
\sum_{t \in T} {|E| \choose \beta_t}=2 \cdot {21 \choose 2}=420.
\end{align}
Next, we calculate the bound $\sum_{t \in T} \big| R_t(\beta_t) \big|$ on the field size $q$ in Proposition~\ref{original-bound_field_size_LNEC_codes}. By~\eqref{R_t} and $\beta_{t_1}=2$, we obtain that
\begin{align*}
R_{t_1}(2)=&\Big\{
     \{ e_{1}, e_{2} \}, \{ e_{1}, e_{3} \}, \{ e_{1}, e_{4} \}, \{ e_{1}, e_{5} \},
     \{ e_{1}, e_{6} \}, \{ e_{1}, e_{7} \}, \{ e_{1}, e_{8} \}, \{ e_{1}, e_{10} \}, \\
    &\{ e_{1}, e_{12} \}, \{ e_{1}, e_{13} \}, \{ e_{1}, e_{14} \}, \{ e_{1}, e_{16} \},
     \{ e_{1}, e_{17} \}, \{ e_{1}, e_{18} \}, \{ e_{1}, e_{20} \}, \{ e_{2}, e_{3} \},\\
    &\{ e_{2}, e_{4} \}, \{ e_{2}, e_{5} \}, \{ e_{2}, e_{6} \}, \{ e_{2}, e_{10} \},
     \{ e_{2}, e_{12} \}, \{ e_{2}, e_{13} \}, \{ e_{2}, e_{14} \}, \{ e_{2}, e_{17} \},\\
    &\{ e_{2}, e_{20} \}, \{ e_{3}, e_{4} \}, \{ e_{3}, e_{5} \}, \{ e_{3}, e_{6} \},
     \{ e_{3}, e_{7} \}, \{ e_{3}, e_{8} \}, \{ e_{3}, e_{12} \}, \{ e_{3}, e_{13} \},\\
    &\{ e_{3}, e_{14} \}, \{ e_{3}, e_{16} \}, \{ e_{3}, e_{17} \}, \{ e_{3}, e_{18} \},
     \{ e_{3}, e_{20} \}, \{ e_{4}, e_{5} \}, \{ e_{4}, e_{6} \}, \{ e_{4}, e_{7} \},\\
    &\{ e_{4}, e_{8} \}, \{ e_{4}, e_{10} \}, \{ e_{4}, e_{12} \}, \{ e_{4}, e_{13} \},
     \{ e_{4}, e_{14} \}, \{ e_{4}, e_{16} \}, \{ e_{4}, e_{17} \}, \{ e_{4}, e_{18} \},\\
    &\{ e_{4}, e_{20} \}, \{ e_{5}, e_{6} \}, \{ e_{5}, e_{7} \}, \{ e_{5}, e_{8} \},
     \{ e_{5}, e_{10} \}, \{ e_{5}, e_{12} \}, \{ e_{5}, e_{16} \}, \{ e_{5}, e_{18} \},\\
    &\{ e_{6}, e_{7} \}, \{ e_{6}, e_{8} \}, \{ e_{6}, e_{10} \}, \{ e_{6}, e_{12} \},
     \{ e_{6}, e_{13} \}, \{ e_{6}, e_{14} \}, \{ e_{6}, e_{16} \}, \{ e_{6}, e_{17} \},\\
    &\{ e_{6}, e_{18} \}, \{ e_{6}, e_{20} \}, \{ e_{7}, e_{10} \}, \{ e_{7}, e_{12} \},
     \{ e_{7}, e_{13} \}, \{ e_{7}, e_{14} \}, \{ e_{7}, e_{17} \}, \{ e_{7}, e_{20} \},\\
    &\{ e_{8}, e_{10} \}, \{ e_{8}, e_{12} \}, \{ e_{8}, e_{13} \}, \{ e_{8}, e_{14} \},
     \{ e_{8}, e_{17} \}, \{ e_{8}, e_{20} \}, \{ e_{10}, e_{12} \}, \{ e_{10}, e_{13} \},\\
    &\{ e_{10}, e_{14} \}, \{ e_{10}, e_{16} \}, \{ e_{10}, e_{17} \}, \{ e_{10}, e_{18} \},
     \{ e_{10}, e_{20} \}, \{ e_{12}, e_{13} \}, \{ e_{12}, e_{14} \}, \{ e_{12}, e_{16} \},\\
    &\{ e_{12}, e_{17} \}, \{ e_{12}, e_{18} \}, \{ e_{12}, e_{20} \}, \{ e_{13}, e_{16} \},
     \{ e_{13}, e_{18} \}, \{ e_{14}, e_{16} \}, \{ e_{14}, e_{18} \}, \{ e_{16}, e_{17} \},\\
    &\{ e_{16}, e_{20} \}, \{ e_{17}, e_{18} \}, \{ e_{18}, e_{20} \} \Big\}
\end{align*}
with $|R_{t_1}(2)|=99$. By the symmetry of the network $G$, we also have $|R_{t_2}(2)|=99$. So, the bound~\eqref{Fsize-LNEC-GuangFZ} in Proposition~\ref{original-bound_field_size_LNEC_codes} is
\begin{align}\label{GFZ-Bound-EX}
|R_{t_1}(2)|+|R_{t_2}(2)|=198,
\end{align}
which is smaller than $420$ from~\eqref{straightforward-Bound-EX}.

Next, we present the set $\mA_{t_1}(2)$ of all the primary edge subsets for $t_1$ of size $\beta_{t_1}=2$ as follows:
\begin{align*}
\mA_{t_1}(2)=&\Big\{
      \{ e_{1}, e_{4} \}, \{ e_{1}, e_{10} \}, \{ e_{1}, e_{12} \}, \{ e_{1}, e_{20} \},
     \{ e_{6}, e_{10} \}, \{ e_{6}, e_{12} \}, \{ e_{6}, e_{18} \},\\
     &\{ e_{6}, e_{20} \}, \{ e_{10}, e_{12} \}, \{ e_{10}, e_{18} \}, \{ e_{10}, e_{20} \},
     \{ e_{12}, e_{18} \}, \{ e_{12}, e_{20} \}, \{ e_{18}, e_{20} \} \Big\}.
\end{align*}
Then, $|\mA_{t_1}(2)|=14$. We also have $|\mA_{t_2}(2)|=|\mA_{t_1}(2)|=14$. Thus, the improved bound~\eqref{equ-field_size_LNEC} in Theorem~\ref{thm_improved_field_size_LNEC_codes} is
$$|\mA_{t_2}(2)|+|\mA_{t_1}(2)|=28,$$
which is considerably smaller than $198$ from~\eqref{GFZ-Bound-EX}.
\end{example}

On the other hand, by the definition of primary edge subset in the paragraph immediately above Example~\ref{EX_1}, it is not difficult to see that for a sink node $t$, any $\beta_t$ of the $|\In(t)|$ input edges of~$t$ form a size-$\beta_t$ primary edge subset for $t$. We thus immediately obtain a lower bound on the size of $\mA_t(\beta_t)$ as presented in the following corollary.

\begin{cor}\label{thm_lower_bound_on_mA}
For a sink node $t$, let $\beta_t$ be a nonnegative integer not larger than $C_t-\w$. Then
\begin{align*}
|\mA_t(\beta_t)|\geq {|\In(t)| \choose \beta_t}.
\end{align*}
\end{cor}

Continuing from Example~\ref{eg_field-size}, by this corollary, the size $14$ of $\mA_{t_i}(\beta_{t_i})$ is lower bounded by ${|\In({t_i})| \choose \beta_{t_i}}=10$ for $i=1,2$.

Next, we will present an improved upper bound on the minimum required field size for the existence of a rate-$\w$ LNEC MDS code in the following theorem which is a consequence of~Theorem~\ref{thm_improved_field_size_LNEC_codes}. First, we recall that an $\Fq$-valued rate-$\w$ LNEC code $\tC$ is MDS if this code $\tC$ is decodable for $T$ and satisfies the Singleton bound~\eqref{SingletonBound_t} with equality, i.e.,
\begin{align*}
\dim\big(\Phi(t)\big)=\w \quad \text{and} \quad  d_{\min}^{(t)}=C_t-\w+1, ~~ \forall~t\in T.
\end{align*}

\begin{thm}\label{improved_bound}
Let $\mF$ be a finite field of order $q$, where $q$ is a prime power, and $T$ be the set of sink nodes on the network $G$ with $C_t\geq \w$, $\forall~t\in T$. There exists an $\Fq$-valued rate-$\w$ LNEC MDS code on $G$ if the field size $q$ satisfies
\begin{align}\label{Fsize-m-MDS}
q > \sum_{t \in T}\big|\mA_t(\de_t)\big|,
\end{align}
where $\dt_t \triangleq C_t-\w$ is called the redundancy of the sink node $t\in T$.
\end{thm}





The best known upper bound $\sum_{t\in T} \big|R_t(\de_t)\big|$ on the minimum required field size for the existence of a rate-$\w$ LNEC MDS code was presented in~\cite{Guang-MDS} (cf.~\cite[Theorem~5]{Guang-MDS}). The bound in Theorem~\ref{improved_bound} improves this bound and the improvement is in general significant. In fact, the LNEC code considered in Example~\ref{eg_field-size} is MDS and we have seen that the improvement is significant. Furthermore, similar to Corollary~\ref{thm_lower_bound_on_mA}, a lower bound on the size of $\mA_t(\de_t)$ is given as follows.


\begin{cor}\label{cor_lower_bound_on_mA}
For a sink node $t$, the size of $\mA_t(\de_t)$ is lower bounded by ${|\In(t)| \choose \de_t}$, i.e.,
\begin{align*}
|\mA_t(\de_t)|\geq {|\In(t)| \choose \de_t}.
\end{align*}
\end{cor}

We recall the discussion immediately above Example~\ref{eg_field-size}. Together with the fact that $|\In(t)|=|E|$ for any network as depicted in Fig.~\ref{Fig_classical_coding_model}, the discussion shows that the lower bound in Corollary~\ref{cor_lower_bound_on_mA} is tight, i.e.,
\begin{align*}
|\mA_t(\de_t)| = {|\In(t)| \choose \de_t}, \quad \forall~\de_t \leq |\In(t)|.
\end{align*}
Further, since network error correction coding over such a network depicted in Fig.~\ref{Fig_classical_coding_model} can be regarded as the model of classical coding theory, $|\mA_t(\de_t)| = {|\In(t)| \choose \de_t}$ is an upper bound on the minimum required field size for the existence of an $\big[\,|\In(t)|,~ |\In(t)|-\de_t\,\big]$ linear MDS code, where $|\In(t)|$ and $|\In(t)|-\de_t$ are the length and dimension of the code, respectively. In general, linear MDS codes with field size smaller than this bound exist. For example, let $|E|=n$ and $\de_t=n-k$, where $k$ ($k\leq n$) is the designed dimension of the code. Then, there exists an $[n,\, k]$ linear MDS code over a finite field $\Fq$ if $q\geq n-1$. A well-known conjecture on the field size for the existence of linear MDS codes is the following.

\medskip

\indent  \textbf{MDS Conjecture} (\!\!\cite[Chapter~7.4]{Huffman-Pless-Fundamentals-error-correting-codes_book}):~{\it
If there is a nontrivial $[n,\, k]$ linear MDS code over $\Fq$, then $n \leq  q+1$, except when $q$ is even and $k=3$ or $k=q-1$, in which case $n \leq  q+2$.}

\subsection{Efficient Algorithm for Computing the Improved Bound}

In the last subsection, an improved upper bound on the minimum required field size for the existence of LNEC codes is obtained. The bound thus obtained is graph-theoretic, which depends only on the network topology and the required error correction capability of the LNEC code. However, it is not given in a form which is readily computable. Accordingly, we in this subsection will develop an efficient algorithm to compute this bound.

Let $t$ be a sink node on the network $G=(V,E)$ and $r$ be a nonnegative integer not larger than $C_t-\w$. We first develop an efficient algorithm for computing $\mA_t(r)$. An implementation of the algorithm is given in Algorithm~\ref{algo_compt-mA}.

\begin{algorithm}[!htb]

\medskip
\SetAlgoLined
\KwIn{The network $G=(V, E)$,  a sink node $t$ and a nonnegative integer $r$.}
\KwOut{$\mA_t(r)$, the set of all the size-$r$ primary edge subsets for $t$.}
\BlankLine
\Begin{
\nl\label{algo-compt-mA_line-init_mA}
Set $\mA(r)=\emptyset$\;
\nl\label{algo-compt-mA_line-init_mB}
Set $\mB=\big\{ \eta\subseteq E_t:~|\eta|=r \big\}$, where $E_t$ is the set of the edges in $E$ from which $t$ is reachable\;
\tcp*[f]{\rm\footnotesize If there exists a directed path from an edge $e$ to $t$, we say $t$ is reachable from $e$ or $e$ can reach $t$.}

\nl\label{algo-compt-mA_line-while} \While{$\mB \neq \emptyset$}{
\nl\label{algo-compt-mA_line-chose_A}      choose an edge subset $\eta$ in $\mB$\;
\nl\label{algo-compt-mA_line-find_PrimCUT}      find the primary minimum cut $\p$ separating $t$ from $\eta$\;
\tcp*[f]{\rm\footnotesize The primary minimum cut $\p$ separating $t$ from $\eta$ is a primary edge subset for $t$.}

\nl\label{algo-compt-mA_line-eif}      \eIf(\tcp*[f]{\rm\footnotesize Namely, $|\p|<r$.}){ $|\p| \neq r$ }{
\nl\label{algo-compt-mA_line-remove_A}      remove $\eta$ from $\mB$;}
(\tcp*[f]{\rm\footnotesize Namely, $|\p|=r$.}){
\nl\label{algo-compt-mA_line-add_CUT}      add $\p$ to $\mA(r)$\;
\nl\label{algo-compt-mA_line-partition}      partition $E_t$ into two parts $E_{t,\p}$ and $E_{t,\p}^c=E_t\setminus E_{t,\p}$\;
\tcp*[h]{\rm\footnotesize Here, $E_{t,\p}$ is the set of the edges from which $t$ is reachable upon deleting the edges in $\p$.}\newline
\tcp*[h]{\rm\footnotesize Note that $\p\subseteq E_{t,\p}^c$.}\newline
\nl\label{algo-compt-mA_line-for}      \For{ each $\mu\in \mB$ }{
\nl\label{algo-compt-mA_line-if}      \If{ $\mu\subseteq E_{t,\p}^c$ }{
\nl\label{algo-compt-mA_line-remove-B}      remove $\mu$ from $\mB$;}
         } 
}
}
\nl\label{algo-compt-mA_line-return_mA}      Return $\mA(r)$.\newline
\tcp*[f]{\rm\footnotesize After the ``while'' loop, $\mA(r)$ contains all the size-$r$ primary edge subsets for $t$, i.e., $\mA(r)=\mA_t(r)$.}
}
\caption{Algorithm for computing $\mA_t(r)$}
\label{algo_compt-mA}
\end{algorithm}

\bigskip

\noindent\textbf{Algorithm Verification:}

\begin{enumerate}
  \item In Lines~\ref{algo-compt-mA_line-init_mA} and~\ref{algo-compt-mA_line-init_mB}, initialize two sets $\mA(r)$ and $\mB$ to the empty set and the set of all size-$r$ edge subsets of $E_t$, respectively, where $E_t$ denotes the set of edges in $E$ from which $t$ is reachable, i.e., for each $e\in E_t$, there exists a directed path from $e$ to $t$ on the network $G$.\label{step1}
  \item In Lines~\ref{algo-compt-mA_line-chose_A} and~\ref{algo-compt-mA_line-find_PrimCUT}, arbitrarily choose an edge subset $\eta \in \mB$ and find the primary minimum cut separating $t$ from $\eta$, denoted by $\p$. We note that for each edge subset $\eta$, the primary minimum cut separating $t$ from $\eta$ exists and is unique.\label{step2}
  \item We note that
  \begin{align}\label{equ1:verify-algo_compt-mA}
  |\p|=\mincut(\eta,t)\leq |\eta|=r,
  \end{align}
  and then consider two cases below.

      \noindent\textbf{Case 1:}  If $|\p|\neq r$, which implies $|\p|< r$ by~\eqref{equ1:verify-algo_compt-mA}, then the ``if'' statement (Line~\ref{algo-compt-mA_line-remove_A}) is executed. In this case, we readily see that $\p$ is not a size-$r$ primary edge subset for $t$. Then, we remove $\eta$ from $\mB$ and go back to Line~\ref{algo-compt-mA_line-while} for checking whether the updated $\mB$ is empty or not.

      \noindent\textbf{Case 2:}  If $|\p|=r$, which implies that $\p$ is a size-$r$ primary edge subset for $t$, then the ``else'' statement (Lines~\ref{algo-compt-mA_line-add_CUT}--\ref{algo-compt-mA_line-remove-B}) is executed.
      To be specific, in Line~\ref{algo-compt-mA_line-add_CUT}, add this size-$r$ primary edge subset $\p$ to $\mA(r)$. In Line~\ref{algo-compt-mA_line-partition}, partition the edge set $E_t$ into two disjoint subsets: $E_{t,\p}$ and $E_{t,\p}^c\triangleq E_t\setminus E_{t,\p}$, where $E_{t, \p}$ is the set of edges from which $t$ is reachable upon deleting the edges in $\p$. Note that $\p\subseteq E_{t, \p}^c$. Next, for the ``for'' loop (Lines~\ref{algo-compt-mA_line-for}--\ref{algo-compt-mA_line-remove-B}), all the edge subsets in $\mB$ that are subsets of $E_{t,\p}^c$ are removed. By Lemma~\ref{lem_for_prop}, it is not difficult to see that each edge subset $\eta$ in $\mB$, regardless of whether $\mincut(\eta,t)=r$ or $\mincut(\eta,t)<r$, is a subset of $E_{t,\p}^c$ if and only if $\p$ separates $t$ from $\eta$. This immediately implies that after this ``for'' loop, all the edge subsets in $\mB$ from which~$\p$ separates $t$ are removed from $\mB$, and none of the other size-$r$ primary edge subsets are removed from $\mB$. Thus, we see that in each iteration, exactly one size-$r$ primary edge subset for $t$ is added to $\mA(r)$. \label{step3}

 \item Repeat Steps \ref{step2}) and \ref{step3}) above until $\mB$ is empty and output $\mA(r)$ in Line~\ref{algo-compt-mA_line-return_mA}, which is now equal to $\mA_t(r)$.\label{step5}
\end{enumerate}

\bigskip

In Algorithm~\ref{algo_compt-mA}, the two crucial steps are \rmnum{1}) to find the primary minimum cut $\p$ separating $t$ from an edge subset $\eta$ in $\mB$ (Line~\ref{algo-compt-mA_line-find_PrimCUT}), and \rmnum{2}) to partition $E_{t}$ into $E_{t,\p}$ and $E_{t,\p}^c$ (Line~\ref{algo-compt-mA_line-partition}). We first consider the step of partitioning $E_{t}$ into $E_{t,\p}$ and $E_{t,\p}^c$. Toward this end, it suffices to determine the edge set $E_{t,\p}$, i.e., to find all the edges that can reach $t$ upon deleting the edges in $\p$. This can be implemented efficiently by Algorithm~\ref{algo_search} below.

\begin{algorithm}[!htb]

\medskip

\SetAlgoLined
\KwIn{The network $G=(V, E)$ and a primary edge subset $\p$ for $t$.}
\KwOut{$E_{t,\p}$, the set of all the edges that can reach $t$ upon deleting the edges in $\p$.}
\BlankLine

\Begin{
\nl Unmark all nodes in $V$\;
\nl mark sink node $t$\;
\nl set an edge-set $\eSET=\emptyset$\;
\nl set a node-set $\nSET=\{t\}$\;
\nl \While{$\nSET \neq \emptyset$}{
\nl\label{algo_searchh-line-node_v}      select a node $v$ in $\nSET$\;
\nl\label{algo_searchh-line-for}      \For{each node $u$ incident to an edge $(u,v)$ not in $\p$}{
\nl\label{algo_searchh-line-add-edges}     add all parallel edges leading from $u$ to $v$ and not in $\p$ to $\eSET$\;
\nl      \If{$u$ is unmarked}{
\nl\label{algo_searchh-line-mark-u}          mark node $u$\;
\nl\label{algo_searchh-line-add-u}          add node $u$ to $\nSET$\;
} 
} 
\nl\label{algo_searchh-line-remove-v}           delete node $v$ from $\nSET$\;
}
\nl Return $\eSET$.\newline
\tcp*[f]{\rm\footnotesize After the ``while'' loop, $\eSET$ contains all the edges that can reach $t$ upon deleting the edges in $\p$, i.e., $\eSET=E_{t,\p}$.}
         }
\caption{Algorithm for partitioning $E_{t}$ into $E_{t,\p}$ and $E_{t,\p}^c$}
\label{algo_search}
\end{algorithm}

Algorithm~\ref{algo_search} extends from the sink node $t$ and identifies an increasing number of edges that can reach $t$. At any point during the execution of the algorithm, all the nodes in the network can be in one of two states: marked or unmarked. The marked nodes are those from which $t$ is reachable, and the unmarked nodes are those yet to be classified. The edges in the set $\eSET$ at this point have been identified to be those from which $t$ is  reachable. The set $\nSET$ contains marked nodes whose input edges have not been processed. When a node $v\in \nSET$ is selected in Line~\ref{algo_searchh-line-node_v}, all the input edges of $v$ that are not in $\p$ are added to $\eSET$ in the ``for'' loop (Lines~\ref{algo_searchh-line-for}--\ref{algo_searchh-line-add-u}). Since $v\in \nSET$, we see that $v$ is marked and so $t$ is reachable from $v$. This implies that $t$ is reachable from all these input edges and they are added to $\eSET$ in Line~\ref{algo_searchh-line-add-edges}. The node $u$ incident to an edge $(u,v)$ can reach $t$ via node $v$. If $u$ is unmarked, then mark $u$ in Line~\ref{algo_searchh-line-mark-u}. Otherwise, $u$ has already been marked and so $t$ is reachable from $u$. After the ``for'' loop (Lines~\ref{algo_searchh-line-for}--\ref{algo_searchh-line-add-u}), all the input edges of $v$ that are not in $\p$ are added to $\eSET$ and all the nodes $u$ incident to an edge $(u,v)$ are marked. Now, the node $v$ has been processed and is removed from $\nSET$ in Line~\ref{algo_searchh-line-remove-v}. The algorithm terminates when the set of nodes $\nSET$ is empty. At this point, all the nodes that can reach $t$ have been marked and processed, and the edge set $\eSET$ contains all the edges that can reach $t$ upon deleting the edges in $\p$, namely that $\eSET=E_{t,\p}$. Now, we consider the complexity of Algorithm~\ref{algo_search}. We can readily see that the algorithm traverses all the edges in $E_{t,\p}$ exactly once, and thus Algorithm~\ref{algo_search} can find the edge set $E_{t,\p}$ in $\mO(|E_{t,\p}|)$ time.

\bigskip

Next, we consider the other crucial step of finding the primary minimum cut $\p$ separating $t$ from an edge subset $\eta$ in $\mB$. Guang and Yeung \cite{GY-SNC-Reduction} proved that in the augmenting path algorithm \cite{maxflow-Ford-Fulkerson,Elias-Feinstein-Shannon-maxflow} (also see \cite[Chapter 6.5]{Book-NetwFlow} and \cite[Chapter 7.2]{Book-GraphTh-Bondy-Murty}) for finding the maximum flow from the source node~$s$ to a non-source node $t$ on a directed acyclic network, the last step for determining the termination of the algorithm in fact finds the primary minimum cut separating $t$ from $s$. Based on this result, we can develop an efficient algorithm for directly finding the primary minimum cut separating $t$ from~$\eta$, which avoids reversing the network $G$ to $G^\top$ and then finding minimum cuts separating $\eta$ from~$t$ on $G^\top$.

On the network $G$, we first subdivide each edge $e\in \eta$ by creating a node $v_e$ for $e$ and splitting $e$ into two edges $e^1$ and $e^2$ with $\tail(e^1)=\tail(e)$, $\head(e^2)=\head(e)$, and $\head(e^1)=\tail(e^2)=v_e$. Then, we create a new node $v_{\eta}$ and add a new ``super-edge'' with infinite capacity from $v_{\eta}$ to $v_e$ for every node $v_e$, $e\in \eta$. By the definition of a cut separating $t$ from $\eta$ in the first paragraph of Section~\ref{sec:LNEC-revisited}, we can readily see that a cut of finite capacity separating $t$ from $v_\eta$ is a cut separating $t$ from $\eta$ on $G$, and vice versa (where, whenever $e^1$ or $e^2$ appears in the cut, replace it by $e$).
As such, for the purpose of finding the primary minimum cut separating $t$ from~$\eta$ on $G$, we only need to consider algorithms for finding the primary minimum cut separating $t$ from $v_{\eta}$.
Furthermore, for the sake of computational efficiency, in finding the primary minimum cut separating $t$ from~$\eta$ (or equivalently, the primary minimum cut separating $t$ from $v_{\eta}$), it suffices to set the capacities of all the newly added ``super-edges'' $\widehat{e}$ from $v_\eta$ to $v_e$, $e\in \eta$ to one rather than infinity. In fact, the primary minimum cut separating $t$ from $v_{\eta}$ does not contain any newly added super-edge whether its capacity is finite or infinite. To see this, suppose $\p$ is the primary minimum cut separating $t$ from $v_{\eta}$ and assume that it contains a newly added super-edge $\widehat{e}$ from $v_{\eta}$ to $v_e$. Now, we replace $\widehat{e}$ by $e^2$ in $\p$ to form a new edge subset $\p'$, where we recall that $e^2$ is the edge obtained by splitting $e$ with $\tail(e^2)=v_{\eta}$ and $\head(e^2)=\head(e)$. We can see that $\p' \neq \p$ and $\p'$ separates $t$ from $\p$. Thus $\p'$ also separates $t$ from $v_{\eta}$. This contradicts the assumption that $\p$ is the primary minimum cut separating $t$ from $v_{\eta}$.

Let $G=(V, E)$ be a directed acyclic network with a sink node $t$ and a non-sink node $n$. Denote by $C_{n,t}$ the minimum cut capacity separating $t$ from $n$, i.e., $C_{n,t}=\mincut(n,t)$. By the max-flow min-cut theorem \cite{maxflow-Ford-Fulkerson,Elias-Feinstein-Shannon-maxflow}, the value $v(\digamma)$ of a maximum flow $\digamma$ from $n$ to $t$ is equal to the minimum cut capacity $C_{n,t}$, i.e., $v(\digamma)=C_{n,t}$. Since all the edges in the network $G$ have unit-capacity, $C_{n,t}$ is a positive integer and the maximum flow $\digamma$ can be decomposed into $C_{n,t}$ edge-disjoint paths from $n$ to $t$. Such $C_{n,t}$ edge-disjoint paths can be found in polynomial time in $|E|$ \cite{Book-NetwFlow, Book-GraphTh-Bondy-Murty}.
Algorithm~\ref{algo_finding-prmy-mincut} below is an implementation of the algorithm for finding the primary minimum cut separating $t$ from $n$.

\begin{algorithm}[!htb]
\SetAlgoLined
\smallskip
\KwIn{The network $G=(V, E)$ with a maximal flow $\digamma$ from a node $n$ to the sink node $t$ ($n\neq t$). For every edge $e$ in the corresponding $C_{n,t}$ ($\triangleq \mincut(n,t)$) edge-disjoint paths, the flow value is equal to $1$, i.e., $\digamma(e)=1$; otherwise, the flow value is equal to $0$, i.e., $\digamma(e)=0$.}
\KwOut{The primary minimum cut separating $t$ from $n$.}
\BlankLine
\Begin{
\nl Set $S=\{ t \}$\;
\nl \For{ each node $v\in S$ }{
\nl\label{algo_finding-prmy-mincut_Line-if}    \If{ $\exists$ a node $u\in V \setminus S$ s.t. either $\exists$ a reverse edge $e\in E_t$ from $u$ to $v$ s.t. $\digamma(e)=0$ or $\exists$ a forward edge $e\in E_t$ from $v$ to $u$ s.t. $\digamma(e)=1$
}{
\nl      replace $S$ by $S \bigcup \{u\}$.}}
\nl Return $\p=\big\{e:~\tail(e)\in V \setminus S \text{ and } \head(e)\in S \big\}$.
       }
\caption{Algorithm for finding the primary minimum cut separating $t$ from another node $n$}
\label{algo_finding-prmy-mincut}
\end{algorithm}

\begin{example}

\begin{figure}[!t]
\centering
\begin{tikzpicture}
[->,>=stealth',shorten >=1pt,auto,node distance=2.52cm, thick]
  \tikzstyle{every state}=[fill=none,draw=black,text=black,minimum size=6pt,inner sep=0pt]
  \node[state]         (s)[label=above:$s$]                 {};
  \node[state]         (v)[right of=s, xshift=0mm, yshift=-2mm, label=above:$v_{\eta}$]   {};
  \node[state]         (i_2)[below left of=s, xshift=0mm, yshift=-5mm, label=right:$i_2$]   {};
  \node[state]         (v_2)[below left of=s, xshift=7.8mm, yshift=5mm, label={[label distance=-2mm]-10:$v_{e_2}$}]   {};
  \node[state]         (i_4)[below right of=s, xshift=0mm, yshift=-5mm, label=right:$i_4$] {};
  \node[state]         (v_4)[below right of=s, xshift=-7.8mm, yshift=5mm, label={[label distance=-8mm]45:$v_{e_4}$}] {};
  \node[state]         (i_3)[below of=s, yshift=0mm, label=right:$i_3$] {};
  \node[state]         (i_1)[left of=i_2, xshift=0mm, yshift=-8mm, label=left:$i_1$] {};
  \node[state]         (i_5)[right of=i_4, xshift=0mm, yshift=-8mm, label=right:$i_5$] {};
  \node[state]         (i_6)[below right of=i_1, yshift=10mm, label=right:$i_6$] {};
  \node[state]         (i_7)[below of=i_6, yshift=5mm, label=left:$i_7$] {};
  \node[state]         (i_8)[below left of=i_5, yshift=10mm, label=left:$i_8$] {};
  \node[state]         (i_9)[below of=i_8, yshift=5mm, label=right:$i_9$] {};
  \node[state]         (t_1)[below of=i_7, yshift=5mm, label=below:$t_1$] {};
  \node[state]         (t_2)[below of=i_9, yshift=5mm, label=below:$t_2$] {};
\path
(s) edge           node[left=1mm] {$e_1$} (i_1)

    edge           node[pos=0.5, right=-1mm] {$e_2^1$} (v_2)

    edge           node[pos=0.4, left=-2mm] {$e_4^1$} (v_4)

    edge           node[pos=0.7, right=-1mm] {$e_3$} (i_3)
    edge           node[right=1mm] {$e_5$} (i_5)

(v)   edge[ultra thick]          node[pos=0.3, above=0mm] {$\widehat{e}_2$} (v_2)

      edge[ultra thick]           node[pos=0.3, below=0mm] {$\widehat{e}_4$} (v_4)

(v_2) edge[ultra thick]           node[pos=0.6, right=0mm] {$e_2^2$} (i_2)

(v_4) edge[ultra thick]           node[pos=0.6, left=0mm] {$e_4^2$} (i_4)

(i_1) edge         node[left=-1mm]{$e_6$}(t_1)
      edge       node[pos=0.5, right=-0.5mm]{$e_7$}(i_6)
(i_2) edge[ultra thick]       node[pos=0.5, left=-0.5mm]{$e_8$}(i_6)
(i_3) edge       node[pos=0.4, left=-0.5mm]{$e_{10}$}(t_1)
(i_4) edge       node[pos=0.5, right=-0.5mm]{$e_{12}$}(t_1)
      edge[ultra thick]       node[pos=0.5, right=-0.5mm]{$e_{13}$}(i_8)
(i_5) edge       node[pos=0.3, left=0.5mm]{$e_{14}$}(i_8)
(i_6) edge[ultra thick]       node[pos=0.5, right=-1mm]{$e_{16}$}(i_7)
(i_8) edge[ultra thick]       node[pos=0.5, left=-1mm]{$e_{17}$}(i_9)
(i_7) edge[ultra thick]       node[pos=0.3, right=-1.2mm]{$e_{18}$}(t_1)
(i_9) edge[ultra thick]       node[pos=0.9, right=3mm]{$e_{20}$}(t_1);
\end{tikzpicture}
\caption{The network $G_{t_1,\eta}$.}
\label{Fig_network_pri-mincut}
\end{figure}
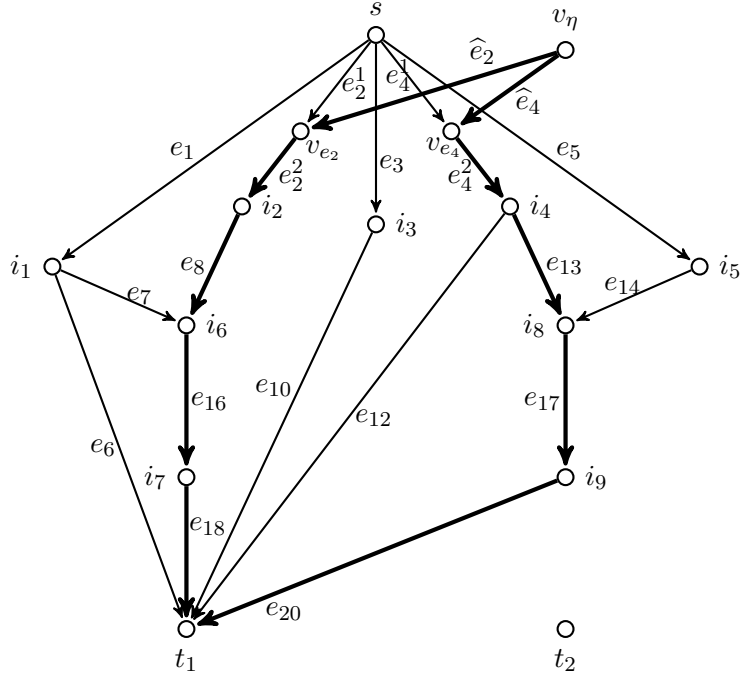

We continue to consider the network $G=(V,E)$ depicted in Fig.~\ref{Fig_network}. In this example, we will illustrate Algorithm~\ref{algo_finding-prmy-mincut} that finds the primary minimum cut separating the sink node $t_1$ from the edge subset $\eta=\{e_2,e_4\}$. Let $G_{t_1,\eta}$ be the network modified from $G$ as illustrated in Fig.~\ref{Fig_network_pri-mincut}. Specifically, from the network $G$, we delete the edges not connected to $t_1$ (i.e., the edges not in $E_{t_1}$); subdivide $e_2$ into two edges $e_2^1$ and $e_2^2$ connected by a newly created node $v_{e_2}$ and subdivide $e_4$ into two edges $e_4^1$ and $e_4^2$ connected by a newly created node $v_{e_4}$; and create a node $v_{\eta}$ with two unit-capacity output edges $\widehat{e}_2$ and $\widehat{e}_4$ leading from $v_{\eta}$ to $v_{e_2}$ and from $v_{\eta}$ to $v_{e_4}$, respectively. Further, a maximum flow $\digamma$ from $v_{\eta}$ to $t_1$ is depicted in Fig.~\ref{Fig_network_pri-mincut}, where all the edges with flow value $1$ are marked in thick lines. In the following, we illustrate Algorithm~\ref{algo_finding-prmy-mincut} that outputs the primary minimum cut separating $t_1$ from $v_{\eta}$ in $G_{t_1,\eta}$, from which we can immediately obtain the primary minimum cut separating $t_1$ from $\eta$ in $G$.
\begin{itemize}
  \item Algorithm~\ref{algo_finding-prmy-mincut} starts with the sink node $t_1$. First, we see that $e_6=(i_1, t_1)$, $e_{10}=(i_3, t_1)$ and $e_{12}=(i_4, t_1)$ are 3 reverse edges incident to $t_1$ with flow value $0$. Thus, the condition of the ``if'' statement in Line~\ref{algo_finding-prmy-mincut_Line-if} is satisfied. We further see that $e_{18}=(i_7, t_1)$ and $e_{20}=(i_9, t_1)$ are 2 reverse edges incident to $t_1$ with flow value $1$, which do not satisfy the condition of the ``if'' statement in Line~\ref{algo_finding-prmy-mincut_Line-if}. Hence, update $S=\{ t_1 \}$ to $\{ t_1, i_1, i_3, i_4 \}$.

  \item We then consider the node $i_1\in S$. The edge $e_1=(s,i_{1})$ with $s\in V\setminus S$ and $i_1\in S$ is a reverse edge with flow value $0$ and thus the condition of the ``if'' statement in Line~\ref{algo_finding-prmy-mincut_Line-if} is satisfied. The edge $e_7=(i_1, i_{6})$ is a forward edge from $i_1$ to $i_{6}$ with $i_6\in V\setminus S$ and $\digamma(e_7)=0$. So the condition of the ``if'' statement in Line~\ref{algo_finding-prmy-mincut_Line-if} is not satisfied. Then, update $S$ to $\{ t_1, s, i_1, i_3, i_4\}$.

  \item For $i_3\in S$, the edge $e_3=(s, i_3)$ is the only edge incident to $i_3$ but the tail node $s$ is already in $S$. So, the condition of the ``if'' statement in Line~\ref{algo_finding-prmy-mincut_Line-if} is not satisfied. Similarly, for $s\in S$, no node in $V\setminus S$ satisfying the condition of the ``if'' statement in Line~\ref{algo_finding-prmy-mincut_Line-if} exists.

  \item For $i_4\in S$, the edge $e_{13}=(i_4,i_{8})$ is a forward edge from $i_4\in S$ to $i_8 \in V\setminus S$ with flow value $1$, which satisfies the condition of the ``if'' statement in Line~\ref{algo_finding-prmy-mincut_Line-if}. Then, update $S$ to $\{ t_1, s, i_1, i_3, i_4, i_8\}$.

  \item For $i_8\in S$, the edge $e_{14}=(i_5,i_8)$ is a reverse edge from $i_5\in V\setminus S$ to $i_8\in S$ with flow value $0$, and the edge $e_{17}=(i_8,i_9)$ is a forward edge from $i_8\in S$ to $i_9 \in V\setminus S$ with flow value $1$. Thus, both $i_5$ and $i_9$ satisfy the condition of the ``if'' statement in Line~\ref{algo_finding-prmy-mincut_Line-if}. Then, update update $S$ to $\{ t_1, s, i_1, i_3, i_4, i_5, i_8, i_9 \}$.

  \item Now, we see that no new node in $V\setminus S$ satisfying the condition of the ``if'' statement in Line~\ref{algo_finding-prmy-mincut_Line-if} exists. Algorithm~\ref{algo_finding-prmy-mincut} terminates and returns the edge set $\p$ below:
      \begin{align*}
      \p=\big\{e:~\tail(e)\in V\setminus S \quad  \text{and}\quad \head(e)\in S \big\}=\big\{ \widehat{e}_4^2=(v_{e_4}, i_4),e_{18}=(i_7, t_1)\big\}.
      \end{align*}
      We readily see that $\p$ is the primary minimum cut separating $t_1$ from $v_{\eta}$ on $G_{t_1,\eta}$.
\end{itemize}

By the definition of a cut separating a node from an edge subset in Section~\ref{sec:enhanced_capability}, the edge subset $\big\{ e_4, e_{18} \big\}$ is the primary minimum cut separating $t_1$ from $\eta$ on $G$.
\end{example}

The computational complexity of Algorithm~\ref{algo_finding-prmy-mincut} is at most $\mO(|E_t|)$ since in the algorithm, each edge in $E_t$ is examined at most once. If we use the augmenting path algorithm to find $C_{n,t}$ edge-disjoint paths from $n$ to $t$, then Algorithm~\ref{algo_finding-prmy-mincut} is already incorporated, and the total complexity for finding the primary minimum cut separating $t$ from $n$ is at most $\mO(C_{n,t}\cdot|E_t|)$, because the path augmentation approach requires at most $\mO(|E_t|)$ time as mentioned and the number of the path augmentations is upper bounded by the minimum cut capacity~$C_{n,t}$.


\bigskip

Now, we can analyze the total complexity of Algorithm~\ref{algo_compt-mA} for computing $\mA_t(r)$. By combining the foregoing discussions, we see that the complexity of Algorithm~\ref{algo_compt-mA} is linear time in $|E_t|$. This is elaborated as follows: \rmnum{1}) The complexity for finding the primary minimum cut $\p$ separating $t$ from an edge subset $\eta$ (Line~\ref{algo-compt-mA_line-find_PrimCUT} in Algorithm~\ref{algo_compt-mA}) is at most $\mO(|E_t|)$; \rmnum{2}) The complexity for partitioning $E_t$ into two parts $E_{t,\p}$ and $E_{t,\p}^c$ (Line~\ref{algo-compt-mA_line-partition} in Algorithm~\ref{algo_compt-mA}) is at most $\mO(|E_{t,\p}|)$, not larger than $\mO(|E_t|)$; \rmnum{3}) Removing all the edge subsets in $\mB$ that are subsets of $E_{t,\p}^c$ (Lines~\ref{algo-compt-mA_line-for}--\ref{algo-compt-mA_line-remove-B} in Algorithm~\ref{algo_compt-mA}) can be implemented by creating an appropriate data structure to avoid computational complexity; \rmnum{4}) The ``while'' loop (Line~\ref{algo-compt-mA_line-while} in Algorithm~\ref{algo_compt-mA}) is executed $|\mA_t(r)|$ times.\footnote{Here, it suffices to consider edge subsets $\eta\in \mB$ with $\mincut(\eta,t)=r$.} So the complexity of Algorithm~\ref{algo_compt-mA} is at most $\mO\big(|\mA_t(r)|\cdot |E_t|)$, that is linear time in $|E_t|$.



\section{Conclusion}\label{sec:concl}

In this paper, we revisited and explored the framework of LNEC coding and network error correction on a network of which the topology is known. Then, we showed that the two well-known LNEC approaches in the literature are in fact equivalent. Further, we enhanced the characterization of error correction capability of LNEC codes in terms of the minimum distances at the sink nodes by developing a graph-theoretic approach. Based on this result, the computational complexities for decoding and code construction can be significantly reduced.

In LNEC coding, the minimum required field size for the existence of LNEC codes, in particular LNEC MDS codes, is an open problem not only of theoretical interest but also of practical importance. However, the existing upper bounds on the minimum required field size for the existence of LNEC (MDS) codes are typically too large for implementation. In this paper, we proved an improved upper bound on the minimum required field size, which shows that the required field size for the existence of LNEC (MDS) codes can be reduced significantly in general. This new bound only depends on the network topology and the requirement of error correction capability but not on a specific code construction. However, it is not given in an explicit form. Thus, we developed an efficient algorithm that computes the upper bound in a linear time of the number of edges in the network. In developing the upper bound and the efficient algorithm for computing this bound, various graph-theoretic concepts are introduced. These concepts appear to be of fundamental interest in graph theory and they may have further applications in graph theory and beyond.




\numberwithin{thm}{section}
\appendices

\section{Proof of Proposition~\ref{prop_distance}}\label{pf_prop_distance}

The positive definiteness and symmetry are straightforward. To complete the proof, we only need to prove the triangle inequality. Consider three arbitrary vectors $\tby_t$, $\tby_t'$ and $\tby_t''$ in $\Fq^{|\In(t)|}$. Let $$d^{(t)}(\tby_t, \tby_t')=d_1\quad \text{ and }\quad  d^{(t)}(\tby_t', \tby_t'')=d_2.$$
Let $\p_1\subseteq E$ be an edge subset with $|\p_1|=d_1$ such that there exists an error vector $\bz_1\in\p_1$ satisfying
\begin{align}\label{app-equ1}
\tby_t-\tby_t'=\bz_1 \cdot G_t,
\end{align}
and similarly $\p_2\subseteq E$ be an edge subset with $|\p_2|=d_2$ such that there exists an error vector $\bz'\in\p_2$ satisfying
\begin{align}\label{app-equ2}
\tby_t'-\tby_t''=\bz' \cdot G_t.
\end{align}
Combining~\eqref{app-equ1} and~\eqref{app-equ2}, we immediately obtain that
\begin{align}
\tby_t-\tby_t''&=(\tby_t-\tby_t')+(\tby_t'-\tby_t'')\nonumber\\
&=(\bz_1+\bz')\cdot G_t.\label{app-equ3}
\end{align}

Further, we let $\bz_1+\bz' \triangleq (z_e:~e\in E)$ and
$\p\triangleq\{e\in E:~ z_e\neq 0 \}$. Clearly, $\bz_1+\bz'\in \p$ and
\begin{align*}
|\p|\leq |\p_1|+|\p_2|=d_1+d_2.
\end{align*}
Together with the definition in \eqref{distance}, we immediately see that
\begin{align*}
d^{(t)}(\tby_t, \tby_t'') & \leq |\p| \leq d_1+d_2 = d^{(t)}(\tby_t, \tby_t')+d^{(t)}(\tby_t', \tby_t'').
\end{align*}
We thus have proved the triangle inequality and also Proposition~\ref{prop_distance}.



\begin{thebibliography}{1}


\bibitem{Elias-Feinstein-Shannon-maxflow}
P. Elias, A. Feinstein, and C. E. Shannon, ``A note on maximum flow through a network,'' \textit{IRE Trans. Inf. Theory}, col. 2, vol. 4, pp. 117-119, April 1956.

\bibitem{maxflow-Ford-Fulkerson}
L. R. Ford Jr. and D. R. Fulkerson, ``Maximal flow through a network,'' \textit{Canadian Journal of Mathematics}, vol. 8, no. 3, pp. 399-404, 1956.

\bibitem{Ahlswede-Cai-Li-Yeung-2000}
R. Ahlswede, N. Cai, S.-Y. R. Li, and R. W. Yeung, ``Network information
flow,'' \textit{IEEE Trans. Inf. Theory}, vol. 46, no. 4, pp. 1204-1216, Jul. 2000.

\bibitem{Celebiler-Stette-1978}
M. Celebiler, G. Stette, ``On increasing the down-link capacity of a regenerative satellite repeater in point-to-point communications,'' \textit{Proceedings of the IEEE}, vol. 66, no. 1, pp. 98-100,  Jan. 1978.

\bibitem{Zhang-Yeung-1999}
R. W. Yeung and Z. Zhang, ``Distributed source coding for satellite communications,''
\textit{IEEE Trans. Inf. Theory}, vol. 45, no. 4, pp. 1111-1120, May 1999.


\bibitem{Li-Yeung-Cai-2003}
S.-Y. R. Li, R. W. Yeung, and N. Cai, ``Linear network coding,'' \textit{IEEE
Trans. Inf. Theory}, vol. 49, no. 2, pp. 371-381, Jul. 2003.

\bibitem{Koetter-Medard-algebraic}
R. Koetter and M. M\'{e}dard, ``An algebraic approach
to network coding,'' \textit{IEEE/ACM Trans.  Netw.}, vol. 11, no. 5,
pp. 782-795, Oct. 2003.


\bibitem{Zhang-book}
R. W. Yeung, S.-Y. R. Li, N. Cai, and Z. Zhang, ``Network coding theory,''
\textit{Foundations and Trends in Communications and Information Theory}, vol. 2, nos.4 and 5, pp. 241-381, 2005.

\bibitem{Yeung-book}
R.~W.~Yeung, \textit{Information Theory and Network Coding}.\hskip 1em plus 0.5em
  minus 0.4em\relax New York: Springer, 2008.

\bibitem{Fragouli-book}
C. Fragouli and E. Soljanin, ``Network coding fundamentals,''
\textit{Foundations and Trends in Networking}, vol. 2, no.1, pp. 1-133, 2007.

\bibitem{Fragouli-book-app}
C. Fragouli and E. Soljanin, ``Network coding applications,''
\textit{Foundations and Trends in Networking}, vol. 2, no.2, pp. 135-269, 2007.

\bibitem{Ho-book}
T. Ho and D. S. Lun, \textit{Network Coding: An Introduction}. Cambridge, U.K.: Cambridge Univ. Press, 2008.

\bibitem{Yeung-Cai-coorrect}
N. Cai and R. W. Yeung, ``Network coding and error correction,'' \textit{in
Proc. IEEE Information Theory Workshop 2002}, Bangalore, India, Oct. 2002, pp. 119-122.

\bibitem{Yeung-Cai-correct-1}
R. W. Yeung and N. Cai, ``Network error correction, part \Rmnum{1}: Basic
concepts and upper bounds,'' \textit{Communications in Information and Systems}, vol. 6, pp. 19-36, 2006.

\bibitem{Yeung-Cai-correct-2}
N. Cai and R. W. Yeung, ``Network error correction, part \Rmnum{2}: Lower
bounds,'' \textit{Communications in Information and Systems}, vol. 6, pp. 37-54, 2006.


\bibitem{zhang-correction}
Z. Zhang, ``Linear network error correction codes in packet networks,'' \textit{IEEE Trans. Inf. Theory}, vol. 54, no. 1, pp. 209-218, Jan. 2008.

\bibitem{Yang-refined-Singleton}
S. Yang, R. W. Yeung, and C. K. Ngai, ``Refined Coding Bounds and Code Constructions for Coherent Network Error Correction,'' \textit{IEEE Trans. Inf. Theory}, vol. 57, no. 3, pp. 1409-1424, Mar. 2011.

\bibitem{Koetter-correction}
R. Koetter and F. Kschischang, ``Coding for errors and erasures in
random network coding,'' \textit{IEEE Trans. Inf. Theory}, vol. 54, no. 8, pp. 3579-3591, Aug. 2008.


\bibitem{Silva-K-K-rank-metric-codes}
D. Silva, F. Kschischang, and R. K\"{o}tter, ``A Rank-Metric Approach to Error Control in Random Network Coding,'' \textit{IEEE Trans. Inf. Theory}, vol. 54, no. 9, pp. 3951-3967, Sep. 2008.

\bibitem{Zhang-survey-paper-NEC}
Z. Zhang, ``Theory and applications of network error correction coding,'' \textit{Proceedings of the IEEE}, vol. 99, no. 3, pp. 406-420, March 2011.

\bibitem{Guang-MDS}
X. Guang, F.-W. Fu, and Z. Zhang, ``Construction of Network Error Correction Codes in Packet Networks,''
\textit{IEEE Trans. Inf. Theory}, vol. 59, no. 2, pp. 1030-1047, Feb. 2013.

\bibitem{Guang-Zhang-NECBook}
X.~Guang and Z.~Zhang, \textit{Linear Network Error Correction Coding}.\hskip 1em plus 0.5em  minus 0.4em\relax New York: Springer, 2014.

\bibitem{MacWilliams-Sloane-Theory-error-correting-codes_book}
F.~J.~MacWilliams and N.~J.~A.~Sloane, \textit{The Theory of Error Correcting Codes}. \hskip 1em plus 0.5em
  minus 0.4em\relax Amsterdam, The Netherlands: North-Holland, 1977.

\bibitem{Huffman-Pless-Fundamentals-error-correting-codes_book}
W.~C.~Huffman and V.~Pless, \textit{Fundamentals of Error Correcting Codes}. \hskip 1em plus 0.5em
  minus 0.4em\relax Cambridge, U.K.: Cambridge Univ. Press, 2003.

\bibitem{Shannon48}
C.~E.~Shannon, ``A Mathematical Theory of Communication,'' \textit{Bell Sys. Tech.
Journal}, 27: 379--423, 623--656, 1948.

\bibitem{Yang-weight}
S.~Yang, R.~W.~Yeung, and Z.~Zhang, ``Weight properties of network codes,'' \textit{European Transactions on Telecommunications}, vol.~19, no.~4, pp.~371--383, 2008.

\bibitem{Matsumoto-Singleton}
R. Matsumoto, ``Construction algorithm for network error-correcting codes attaining the singleton bound,'' \textit{IEICE Trans. Fundamentals}, vol.~E90-A, no.~9, pp.~1729--1735, Nov. 2007.

\bibitem{Guang-uni-MDS}
X. Guang, F.-W. Fu, and Z. Zhang, ``Variable-rate linear network error correction MDS codes,'' \textit{IEEE Trans. Inf. Theory}, vol.~62, no.~6, pp.~3147--3164, June 2016.

\bibitem{zhang-random}
H. Balli, X. Yan, and Z. Zhang, ``On randomized linear network codes and their error correction capabilities,'' \textit{IEEE Trans. Inf. Theory}, vol.~55, no.~7, pp.~3148--3160, July 2009.

\bibitem{Cai}
N. Cai ``Valuable messages and random outputs of edges in
linear network coding,''  in \textit{Proc. IEEE Int. Symp. Information Theory}, Seoul, Korea, June 2009, pp. 413--417.

\bibitem{Ho-etc-random}
T. Ho, R. Koetter, M. M\'{e}dard, M. Effros, J. Shi, and D. Karger, ``A random linear network coding approach to multicast,'' \textit{IEEE Trans. Inf. Theory}, vol. 52, no. 10, pp. 4413--4430, Oct. 2006.

\bibitem{KSK-subspace-codes}
A. Khaleghi, D. Silva, and F. R. Kschischang, ``Subspace codes,'' in \textit{Cryptography and Coding 2009}, M. G. Parker Ed., Lecture Notes in Computer Science, vol. 5921, pp. 1-21, 2009.


\bibitem{Silva-Kschischang-Metrics}
D. Silva and F. R. Kschischang, ``On metrics for error correction in network coding,'' \textit{IEEE Trans. Inf. Theory}, vol.~55, no.~12, pp.~5479--5490, Dec. 2009.

\bibitem{Ho-Byzantine}
T. Ho, B. Leong, R. Koetter, M. M\'{e}edard, M. Effros, and D. Karger,
``Byzantine modification detection in multicast networks with random network coding,'' in \textit{IEEE Trans. Inf. Theory}, vol.~54, no.~6, pp. 2798--2803, June 2008




\bibitem{Jaggi-Byzatine-IT08}
S. Jaggi, M. Langberg, S. Katti, T. Ho, D. Katabi, M. Medard, and M. Effros, ``Resilient network coding in the presence of byzantine adversaries,'' \textit{IEEE Trans. Inf. Theory}, vol. 54, no. 6, pp. 2596--2603, Jun. 2008.

\bibitem{Nutman-Langberg-Byzatine-ISIT08}
L.~Nutman and M. Langberg, ``Adversarial models and resilient schemes for network coding,'' in in \textit{Proc. IEEE Int. Symp. Inf. Theory (ISIT)}, Toronto, ON, Canada, July 2008, pp. 171-¨C175.

\bibitem{Kosut-Tong-Tse-Jaggi-Polytope_codes-IT14}
O.~Kosut, L.~Tong, and D. N. C. Tse, ``Polytope codes against adversaries in networks,'' \textit{IEEE Trans. Inf. Theory}, vol.~60, no.~6, pp. 3308-¨C3344, June 2014.




\bibitem{GYF-LEP-SNC}
X. Guang, R. W. Yeung, and F.-W.~Fu, ``Local-Encoding-Preserving Secure Network Coding,'' \textit{IEEE Trans. Inf. Theory}, vol.~66, no.~10, pp. 5965--5994, Oct. 2020.

\bibitem{GY-SNC-Reduction}
X. Guang and R. W. Yeung, ``Alphabet Size Reduction for Secure Network Coding: A Graph Theoretic Approach,'' \textit{IEEE Trans. Inf. Theory}, vol.~64, no.~6, pp. 4513--4529, June 2018.


\bibitem{Guang-SmlFieldSize-SNC-comm-lett}
X. Guang, J. Lu, and F.-W. Fu, ``Small field size for secure network coding, ''\textit{IEEE Commun. Lett.}, vol. 19, no. 3, pp. 375-378, March 2015.






\bibitem{Book-NetwFlow}
R. K. Ahuja, T. L. Magnanti, and J. B. Orlin, \textit{Network Flows: Theory, Algorithms, and Applications}. \hskip 1em plus 0.5em minus 0.4em\relax Englewood Cliffs, NJ: Prentice-Hall, 1993.

\bibitem{Book-GraphTh-Bondy-Murty}
J. A. Bondy and U. S. R. Murty, \textit{Graph Theory}. \hskip 1em plus 0.5em
  minus 0.4em\relax Springer, 2008.

\end{thebibliography}
\end{document}